\documentclass[final]{siamltex}

\title{Velocity integration in a multilayer neural field model of spatial working memory\thanks{This research was supported by NSF grants (DMS-1311755 and DMS-1615737)}} 

\author{
  Daniel B. Poll\thanks{Department of Mathematics, University of Houston, Houston TX 77204
    ({\tt dbpoll@math.uh.edu}).}
  \and
  Zachary P. Kilpatrick\thanks{Department of Applied Mathematics, University of Colorado, Boulder CO 80309 ({\tt zpkilpat@colorado.edu}); \hspace{5mm} Department of Physiology and Biophysics, University of Colorado School of Medicine, Aurora CO, 80045; \hspace{2cm} Department of Mathematics, University of Houston, Houston TX 77204.}
}

\usepackage{cite}
\usepackage{euscript,amsmath,amssymb,amsfonts,graphicx,bm,cite,appendix}
  \usepackage{paralist}
  \usepackage{epstopdf}
  \usepackage{graphics} 
 \usepackage[colorlinks=true]{hyperref}
 \hypersetup{urlcolor=blue, citecolor=red}

\newcommand{\e}{{\rm e}}

\renewcommand{\d}{{\rm d}}
\newcommand{\pd}{\partial}

\newcommand{\mc}{\mathcal }
\newcommand{\ve}{\varepsilon}

\newcommand{\W}{{\mathbf W}}
\newcommand{\bpsi}{\boldsymbol{\psi}}
\renewcommand{\v}{{\mathbf v}}
\newcommand{\bPhi}{\boldsymbol{\Phi}}
\newcommand{\bvphi}{\boldsymbol{\varphi}}
\newcommand{\z}{{\mathbf z}}
\newcommand{\balpha}{\boldsymbol{\alpha}}
\newcommand{\bbeta}{\boldsymbol{\beta}}

\begin{document}

\maketitle
\newcommand{\slugmaster}{%
\slugger{MMedia}{xxxx}{xx}{x}}

\begin{abstract}
We analyze a multilayer neural field model of spatial working memory, focusing on the impact of interlaminar connectivity, spatial heterogeneity, and velocity inputs. Models of spatial working memory typically employ networks that generate persistent activity via a combination of local excitation and lateral inhibition. Our model is comprised of a multilayer set of equations that describes connectivity between neurons in the same and different layers using an integral term. The kernel of this integral term then captures the impact of different interlaminar connection strengths, spatial heterogeneity, and velocity input. We begin our analysis by focusing on how interlaminar connectivity shapes the form and stability of (persistent) bump attractor solutions to the model. Subsequently, we derive a low-dimensional approximation that describes how spatial heterogeneity, velocity input, and noise combine to determine the position of bump solutions. The main impact of spatial heterogeneity is to break the translation symmetry of the network, so bumps prefer to reside at one of a finite number of local attractors in the domain. With the reduced model in hand, we can then approximate the dynamics of the bump position using a continuous time Markov chain model that describes bump motion between local attractors. While heterogeneity reduces the effective diffusion of the bumps, it also disrupts the processing of velocity inputs by slowing the velocity-induced propagation of bumps. However, we demonstrate that noise can play a constructive role by promoting bump motion transitions, restoring a mean bump velocity that is close to the input velocity.
\end{abstract}



\section{Introduction}
\label{sec:intro}

Spatial working memory tasks test the brain's ability to encode information for short periods of time~\cite{funahashi89,pesaran02}. A subject's performance during such tasks can be paired with brain recordings to help determine how neural activity patterns represent memory during a trial~\cite{goldmanrakic95}. In general, working memory involves the retention of information for time periods lasting a few seconds~\cite{baddeley03}. More specifically, spatial working memory involves the short term storage of a spatial variable, such as idiothetic location~\cite{buzsaki13} or a location on a visual display~\cite{durstewitz00}. A well tested theory of spatial information storage on short timescales involves the generation of persistent activity that encodes input during the retention interval~\cite{wang99}. Network models of this activity typically involve local excitation and broader inhibition, producing localized activity packets referred to as {\em bump attractors}~\cite{compte00,laing01}. These models have recently been validated using recordings from oculomotor delayed-response tasks in monkeys~\cite{wimmer14} and from grid cell networks of freely moving rats~\cite{yoon13}. This suggests that studying network mechanisms for generating reliable neural activity dynamics can provide insight into how the brain robustly performs spatial working memory tasks.

In addition to the short term storage of location, several networks of the brain can integrate velocity signals to update a remembered position~\cite{moser08}. Angular velocity of the head is used by the vestibular system to update memory of heading direction~\cite{taube07}. Furthermore, intracellular recordings from goldfish demonstrate that eye position can be tracked by neural circuits that integrate saccade velocity to update memory of eye orientation~\cite{aksay01}. Velocity integration has also been identified in place cell and grid cell networks, which track an animal's idiothetic location~\cite{wills05,hafting05,geva15}. While these networks each possess distinct circuit mechanisms for integrating and storing information, the general dynamics of their stored position variables tends to be similar~\cite{mcnaughton06}. Neuronal networks that support a continuous (line) or approximately continuous (chain) attractor of solutions constitute a unifying framework for modeling these different systems~\cite{knierim12}. One can then consider the effect of noise, network architecture, or erroneous inputs on the accuracy of position memory~\cite{zhang96,renart03,burak09}.

One important feature of spatial working memory, often overlooked in models, is its distributed nature~\cite{haxby00}. Most models focus on the dynamics of persistent activity representing position memory in a single-layer network~\cite{zhang96,compte00,laing02}. However, extensive evidence demonstrates working memory for visuo-spatial and idiothetic position is represented in several distinct modules in the brain that communicate via long-range connectivity~\cite{curtis06,sargolini06}. There are many possible advantages conferred by such a modular organization of networks underlying spatial memory. One well tested theory notes different network layers can represent position memory on different spatial scales, leading to higher accuracy within small-scale layers and wider range in large-scale layers~\cite{burak08}. Furthermore, the information contained in spatial working memory is often needed to plan motor commands, so it is helpful to distribute this signal across sensory, memory, and motor-control systems~\cite{rowe00}. Another advantage of generating multiple representations of position memory is that it can stabilize the memory through redundancy~\cite{schneidman03}. For instance, coupling between multiple layers of a working memory network can reduce the effects of noise perturbations, as we have shown in previous work~\cite{kilpatrick13b}.

In addition to being distributed, the networks that generate persistent activity underlying spatial working memory also appear to be heterogeneous. For instance, prefrontal cortical networks possess a high degree of variation in their synaptic plasticity properties as well as their cortical wiring~\cite{rao99,wang06}. Furthermore, there is heterogeneity in the way place cells from different hippocampal regions respond to changes in environmental cues~\cite{anderson03,lee04}. Along with such between-region variability, there is local variability in the sequenced reactivations of place cells that mimic the activity patterns that typically occur during active exploration~\cite{pfeiffer15}. In particular, these reactivations are saltatory, rather than smoothly continuous, so activity focuses at a discrete location in the network before rapidly transitioning to a discontiguous location. Such activity suggests that the underlying network supports a series of discrete attractors, rather than a continuous attractor~\cite{brody03}.

Given the spatially distributed and heterogeneous nature of neural circuits encoding spatial working memory, we will analyze tractable models that incorporate these features. We are particularly interested in how the architecture of multilayer networks impacts the quality of the encoded spatial memory. In previous work, we examined networks whose interlaminar connectivity was weak and/or symmetric, ignoring the effects of spatial heterogeneity in constituent layers~\cite{kilpatrick13b,kilpatrick15}. In this work, we will depart from the limit of weak coupling, and derive effective equations for the dynamics of bumps whose positions encode a remembered location. Through the use of linearization and perturbation theory, we can thus determine how both the spatial heterogeneity of individual layers and the coupling between layers impact spatial memory storage. In previous work, we found that spatial heterogeneity can help to stabilize memories of a stationary position~\cite{kilpatrick13c}, but such heterogeneities also disrupt the integration of velocity inputs~\cite{poll16}. Thus, it is important to understand the advantages and drawbacks of heterogeneities, and quantify how they trade off with one another.

We focus on a multilayer neural field model of spatial working memory, with arbitrary coupling between layers and spatial heterogeneity within layers. Furthermore, as we are interested in both the retention of memory and the integration of input, we incorporate a velocity-based modulation to the recurrent connectivity which is non-zero when the network receives a velocity signal~\cite{zhang96}. The stationary bump solutions of this network are analyzed in Section \ref{sec:approx}. Since the effects of velocity input and heterogeneity are presumed to be weak, the stationary bump solutions only depend upon the connectivity between layers. Analyzing the stability of bumps, we can determine the marginally stable modes of these bump solutions which will be susceptible to noise perturbations. Subsequently, we derive a one-dimensional stochastic equation that describes the response of the bump solutions to noise, velocity input, and spatial heterogeneity. With this approximation in hand, we can determine the effective diffusion and velocity of bumps using asymptotic methods, which compare well with numerical simulations of the full model (Section \ref{sec:scb}). Lastly, we analyze more nuanced architectures in Section \ref{sec:ndb}, whose bump solutions possess multiple marginally stable modes. As a result, we find we must derive multi-dimensional stochastic equations to describe their dynamics in response to noise. Our work examines in detail the effects of modular network architecture on the coding of spatial working memory.

\section{Multilayer neural field with spatial heterogeneity}
\label{sec:model}
\begin{figure}
\begin{center} \includegraphics[width=6.5cm]{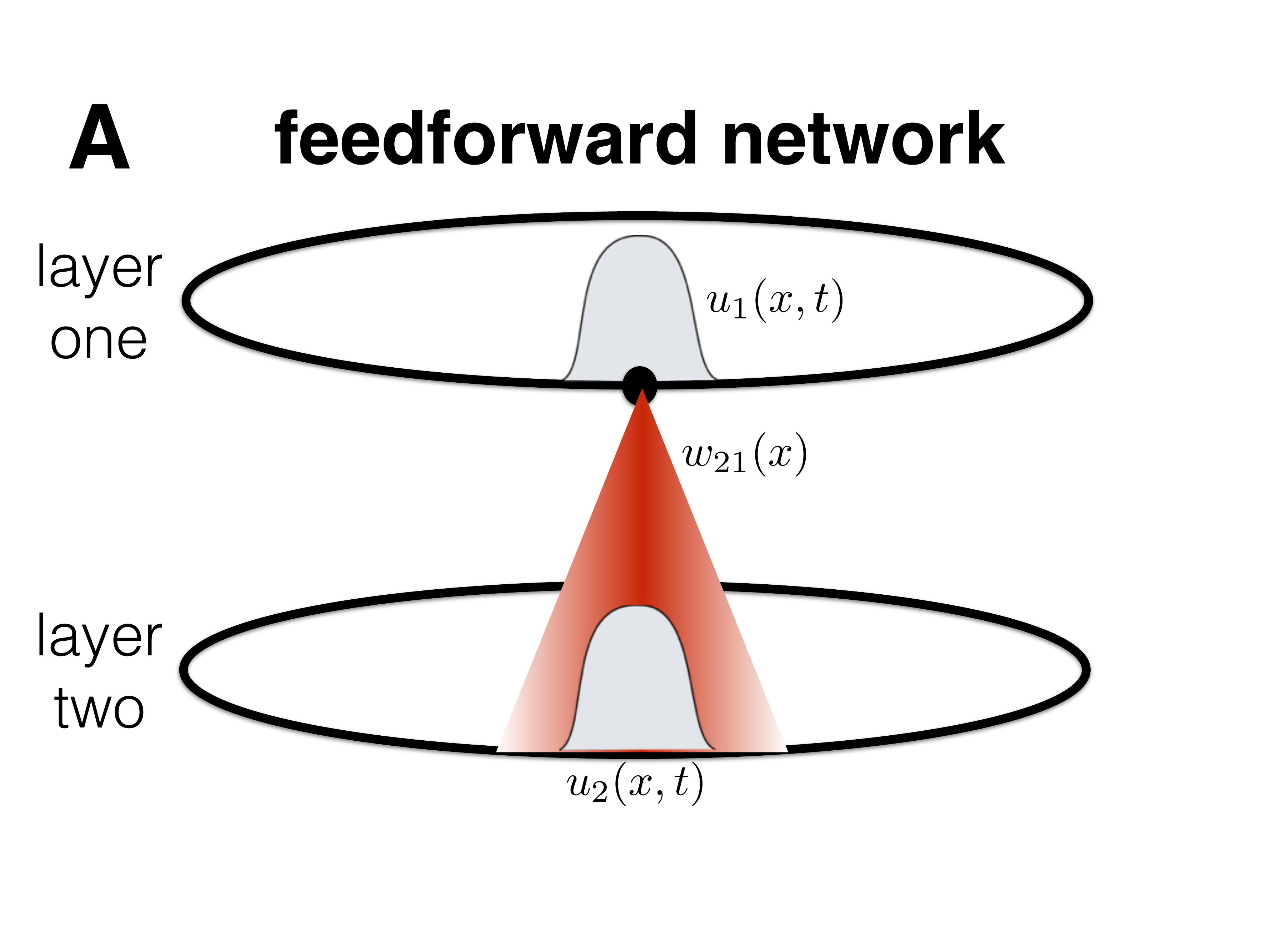} \hspace{7mm} \includegraphics[width=7cm]{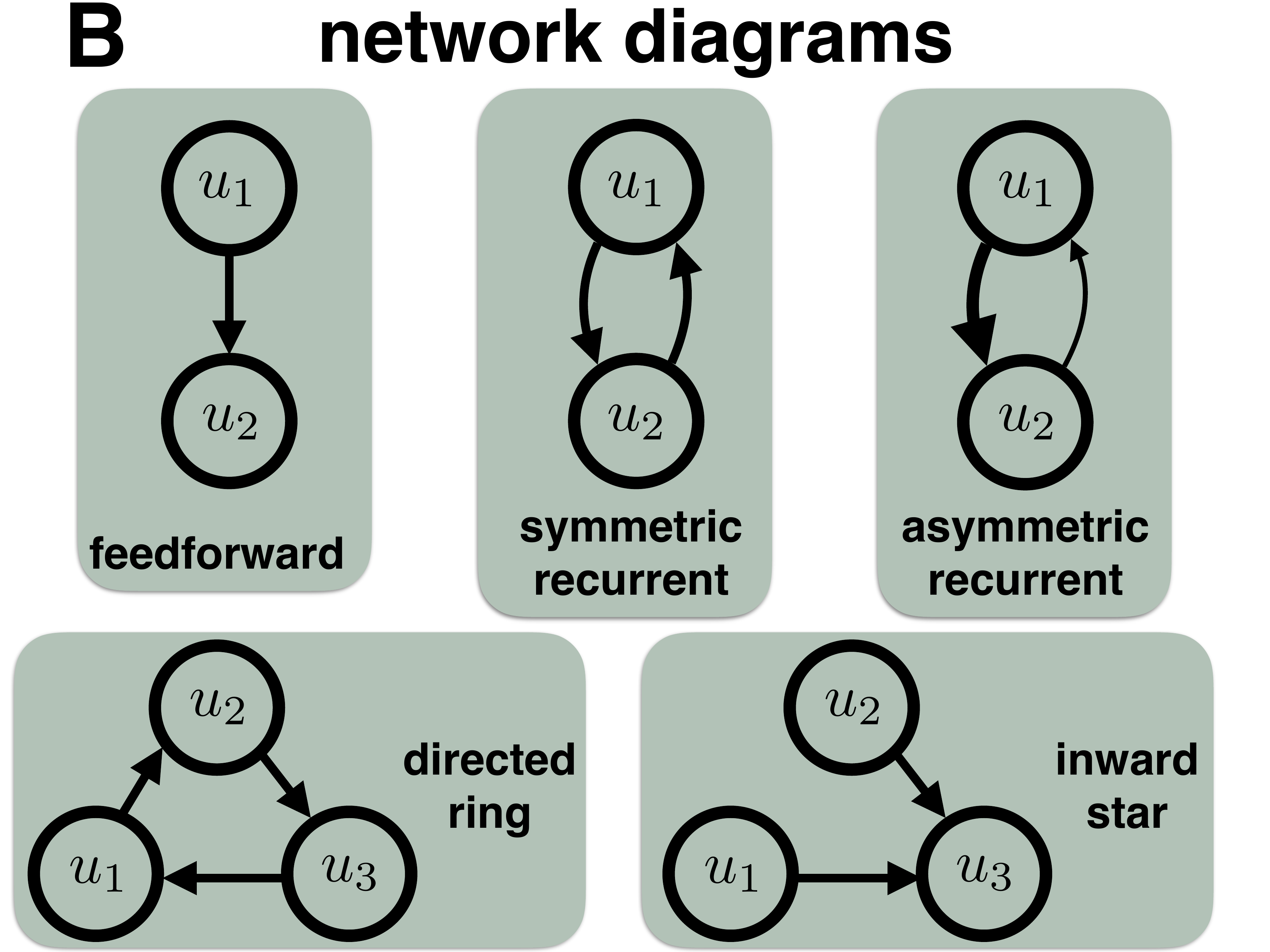} \\[5ex]  \includegraphics[width=7cm]{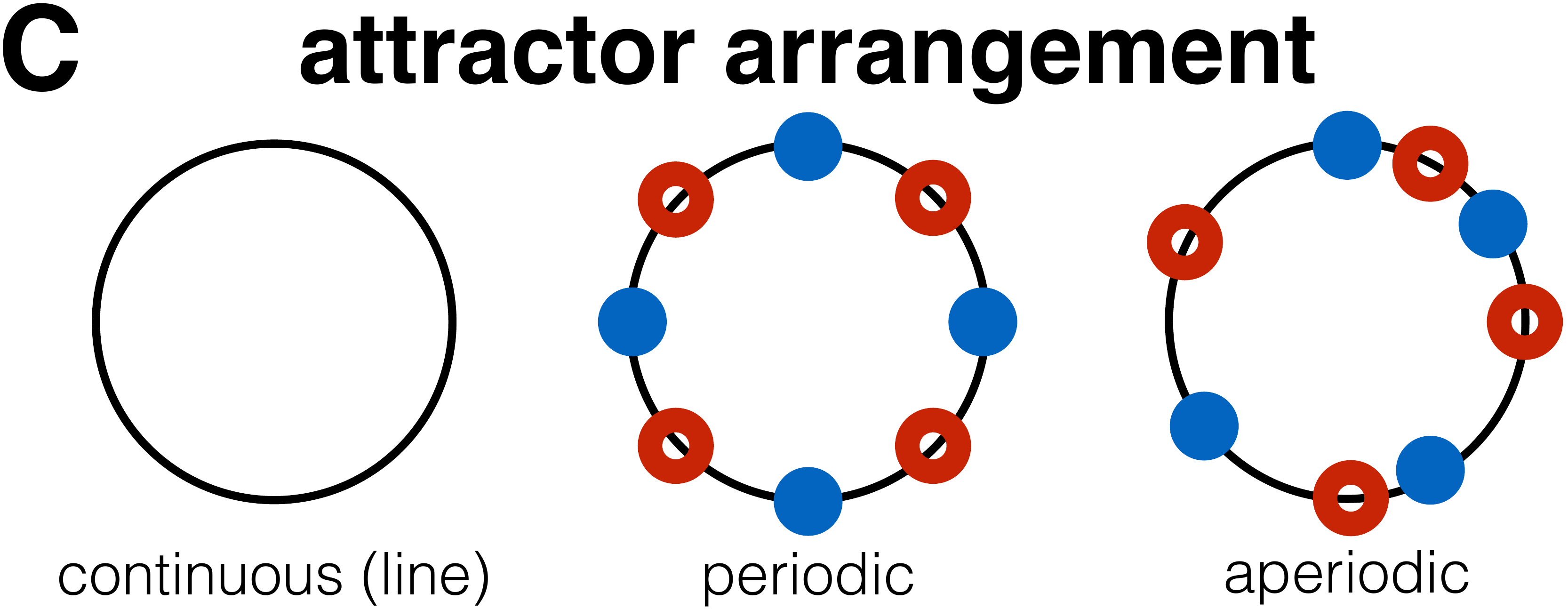} \hspace{7mm} \includegraphics[width=6cm]{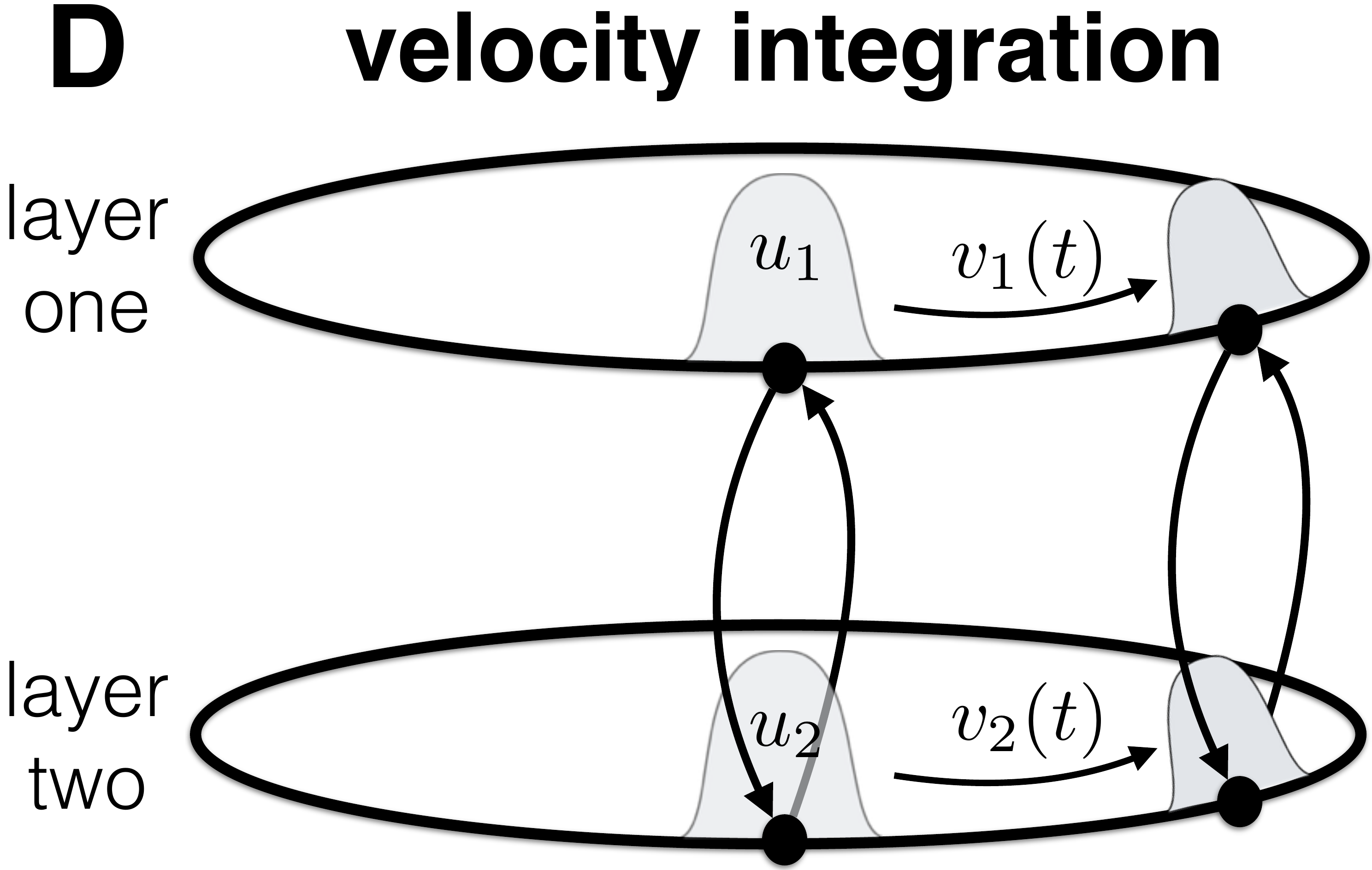}  \end{center}
\caption{Schematic of multilayer network features. ({\bf A}) Purely excitatory long-range interlaminar connectivity is activated by regions of high activity such as the bump attractor solution in layer 1 ($u_1(x,t)$), projecting to similarly tuned locations in layer 2, reinforcing the position of the activity bump there ($u_2(x,t)$). ({\bf B}) Different network topologies as specified by the weight functions ($w_{jk}$, $j \neq k$) are explored in two layer networks (feedforward, symmetric, and asymmetric) as well as three layers (directed ring, inward star). ({\bf C}) Local heterogeneities within each layer introduced into the recurrent weight functions $w_{jj}$, Eq.~(\ref{nfmodel}), generate preferred locations for the bump attractor solutions to the model Eq.~(\ref{rechetero}). We consider a variety of networks, which possess different attractor structures in each of their constituent layers. Continuous attractors possess marginally stable bump solutions at each location around the ring, while chains of discrete attractors possess stable nodes (blue dots) where bumps prefer to reside separated by saddles (red circles). ({\bf D}) Velocity integration via the asymmetric integral term involving $w_{vjk}$ in Eq.~(\ref{nfmodel}) causes bump attractor solutions to move about the domain, incrementing position in proportion to the velocity amplitude.}
\label{fig1}
\end{figure}
Neural field models of persistent activity have been used extensively to understand the relationships between network properties and spatiotemporal activity dynamics~\cite{coombes05,bressloff12}. Stable bump attractors arise as solutions to these models when network connectivity is locally excitatory and broadly inhibitory, and these solutions are translationally invariant when connectivity is also strictly distance-dependent~\cite{amari77,ermentrout98}. However, the incorporation of multiple neural field layers and spatial heterogeneity can break the translation invariance of single network layers, so that bumps have preferred positions within their respective layer~\cite{folias11,folias12,kilpatrick13,kilpatrick13c}. Our analysis focuses on a multilayer neural field model with general connectivity between layers. Spatial heterogeneity within layers, velocity input, and noise are all assumed to be weak (${\mc O}(\ve)$):
\begin{align} 
\label{nfmodel}
\d u_j &= \left[ -u_j + \sum_{k=1}^N \int_{- \pi}^{\pi} w_{jk}(x,y)f(u_k(y,t)) \d y + \ve v(t)  \sum_{k=1}^N w_{vjk}*f(u_k)  \right] \d t + \ve dZ_j,
\end{align}
where $u_j(x,t)$ denotes the average neural synaptic input at location $x \in [- \pi, \pi]$ at time $t$ in network layer $j \in \{ 1, 2, ... , N\}$, and $w_{vjk}*f(u_k) = \int_{- \pi}^{\pi} w_{vjk}(x-y) f(u_k(y,t)) \d y $ is a convolution. Note that we have restricted the spatial domain to be one-dimensional and periodic. There are several experimental examples of spatial working memory which operate on such a domain including oculomotor delayed-response tasks for visual memory~\cite{funahashi89,wimmer14} as well as spatial navigation along linear tracks~\cite{battaglia04,yoon16}. While we suspect that several of our findings extend to two-dimensional spatial domains~\cite{poll15}, we reserve such analysis for future work. Recurrent synaptic connectivity within layers is given by the collection of kernels $w_{jj}(x,y)$, and we allow these functions to be spatially heterogeneous, rather than simply distance-dependent. We thus define them as
\begin{align}
w_{jj}(x,y) : = (1 + \ve h_j(y)) w_{jj}(x-y),  \label{rechetero}
\end{align}
where the impact of the heterogeneity $h_j(y)$ is weak ($\ve \ll1 $), and $w_{jj}(x-y)$ is only dependent on the distance $|x-y|$. As opposed to recurrent connectivity, we assume the interlaminar connectivity ($w_{jk}$, $j \neq k$) is homogeneous, so we can always write $w_{jk}(x,y) = w_{jk}(x-y)$. The homogeneous portion of the recurrent connectivity in each layer is locally excitatory and laterally inhibitory: e.g., the unimodal cosine function
\begin{align}
w_{jj}(x-y) = \cos (x-y), \label{wcos}
\end{align}
which we use in some of our computations. Similarly, we will often consider a cosine shaped excitatory weight function for interlaminar connectivity:
\begin{align}
w_{jk}(x-y) = \bar{w}_{jk} ( 1+ \cos (x-y)). \label{winter}
\end{align}
We introduce homogeneous, distance-dependent kernels for the connectivity between layers. This is motivated by recent experimental work demonstrating that several brain areas involved in spatial working memory are reciprocally coupled to one another~\cite{constantinidis04,curtis06}, and these areas all tend to have similar topographically organized delay period activity~\cite{schluppeck06,kastner07,funahashi89}. Thus, we expect that topologically organized connectivity would be re-enforced via Hebbian plasticity rules~\cite{ko11,qi11}. Such connectivity functions tend to generate stationary bump solutions within each layer~\cite{hansel98,laing01b,kilpatrick13,kilpatrick13b}, and we will analyze these solutions in some detail in Section \ref{sec:approx}. Other lateral inhibitory functions, such as sums of multiple cosine modes, will also generate stationary bump solutions but they do not qualitatively alter the dynamics of the system.

Note, the general form of the weight functions $w_{jk}(x)$ allows us to explore a variety of network topologies, and their impact on the dynamics of bump attractors. For example, it is clear that a feedforward network (Fig. \ref{fig1}A) will primarily be governed by the dynamics of the upstream layer. However, the dynamics of bumps in more intricate networks (Fig. \ref{fig1}B) are more nuanced. Applying both linear stability analysis and perturbation theory to bumps in Section \ref{sec:approx}, we can explore the specific impacts of different conformations of $w_{jk}(x)$. Furthermore, we expect the heterogeneities arising in local connectivity Eq.~(\ref{rechetero}) will interact with interlaminar connectivity to shape the overall dynamics of bumps (Fig. \ref{fig1}C).

The impact of neural activity via synaptic connectivity is thus given via the integral terms, where a nonlinearity is applied to the synaptic input variables:
\begin{align*}
f(u) : = \frac{1}{1 + \e^{- \eta (u - \theta)}},
\end{align*}
and such sigmoids are analogous to the types of saturating nonlinearities that arise from mean field analyses of spiking population models~\cite{brunel99,renart04}. For analytical tractability, we often consider the high gain limit ($\eta \to \infty$) in our examples, resulting in the Heaviside nonlinearity~\cite{amari77,coombes05}
\begin{align}
\lim_{\eta \to \infty} f(u) = H(u - \theta) = \left\{ \begin{array}{ll} 1 & : u > \theta, \\ 0 & : u< \theta. \end{array} \right.  \label{fheav}
\end{align}

The effects of velocity inputs are accounted for by the second integral term in Eq.~(\ref{nfmodel}), based on a well tested model of the head direction system~\cite{taube07} as well as spatial navigation models that implement path integration~\cite{samsonovich97,mcnaughton06}. While some of these models use multiple layers to account for different velocity directions~\cite{xie02,burak09}, the essential dynamics are captured by a single-layer with recurrent connections modulated by velocity input~\cite{zhang96,poll16}. Since we are studying motion along a one-dimensional space, the weak ($\ve \ll 1$) velocity input $ \ve v(t)$ to each neural activity layer $u_j(x,t)$ is given by a scalar variable which can be positive (for rightward motion) or negative (for leftward motion) as shown in Fig. \ref{fig1}D. We derive a reduction of the double ring model (one ring for each velocity direction) of velocity integration presented in \cite{xie02} to a single layer for velocity (positive or negative) in the Appendices. The connectivity functions $w_{vjk}(x-y)$ targeting each layer $j$ should be interpreted as interactions that are shaped by an incoming velocity signal to that layer. Essentially, this connectivity introduces asymmetry into the weight functions, which will cause shifts in the position of spatiotemporal solutions. Typically, this weight function is chosen to be of the form $w_{v}(x-y) = -w'(x-y) $, in single layers~\cite{zhang96}. In the absence of any heterogeneity, such a layer will have bumps that propagate at velocity precisely equal to $\ve v(t)$~\cite{poll16}. As shown in the Appendices \ref{dringone} and \ref{dringmulti}, we can extend this previous assumption to incorporate velocity-related connectivity that respects the interlaminar structure of the network, so that
\begin{align}
w_{vjk}(x-y) = -\frac{\d}{\d x} \left[ w_{jk}(x-y) \right].   \label{velwt}
\end{align}
As we demonstrate in Section \ref{dereff}, this results in bump solutions that propagate with velocity $\ve v(t)$.

Dynamic fluctuations are a central feature of neural activity, and they can often serve to corrupt task pertinent signals, creating error in cognitive tasks~\cite{faisal08}. The error in spatial working memory tasks tends to build steadily in time, in ways that suggest the process underlying the memory may evolve according to a continuous time random walk~\cite{wimmer14,bays15}. As there is no evidence of long timescale correlations in the underlying noise process, we are satisfied to model fluctuations in our model using a spatially correlated white noise process:
\begin{equation*}
\d Z_j(x,t) = \int_\Omega \mathcal{F}_j(x-y) \d Y_j(y,t) \d y,
\end{equation*} 
where $\mathcal{F}_j$ is the spatial filter of the noise in layer $j$ and $\d Y_j(x,t)$ is a spatially and temporally white noise increment. We define the mean and covariance of the vector $( \d Z_1, \d Z_2, ... , \d Z_n )$:
\begin{equation}
\langle \d Z_j(x,t) \rangle \equiv  0 \hspace{1cm} \langle \d Z_j(x,t) \d Z_k(y,t) \rangle = C_{jk}(x-y)\delta(t-s) \d t \d s,  \label{noiseeq}
\end{equation}
where $C_{jk}(x-y)$ is the even symmetric spatial correlation term,
and $\delta(t)$ is the Dirac delta function. 

Subsequently, we will analyze the existence and stability of stationary bump solutions to Eq.~(\ref{nfmodel}) in Section \ref{bumpexs}. Since we will perform this analysis under the assumption of spatially homogeneous synaptic weight functions ($h_j(x) \equiv 0$ in Eq.~(\ref{rechetero})), these solutions will be marginally stable to perturbations that shift their position. However, once we incorporate noise, heterogeneity, and velocity inputs in Section \ref{dereff}, we can perturbatively analyze their effects by linearizing about the stationary bump solutions. The low-dimensional stochastic system we derive will allow us to study the impact of multilayer architecture on the processing of velocity inputs in Section \ref{sec:scb}.

\section{Bump attractors in a multilayer neural field} 
\label{sec:approx}
Our analysis begins by constructing stationary bump solutions to Eq.~(\ref{nfmodel}) for an arbitrary number of layers $N$ and even, translationally-symmetric synaptic weight functions $w_{jk}(x-y)$. Note, there are a few recent studies that have examined the existence and stability of stationary bump solutions to multilayer neural fields~\cite{folias11,folias12,kilpatrick13b}. In particular, Folias and Ermentrout studied bifurcations of stationary bumps in a pair of lateral inhibitory neural field equations~\cite{folias12}. They identified solutions in which bumps occupied the same location in each layer (syntopic) as well as different locations (allotopic), and they also demonstrated traveling bumps and oscillatory bumps that emerged from these solutions. However, they did not study the general problem of an arbitrary number of $N$ layers, and their analysis of networks with asymmetric coupling was relatively limited. Since the solutions will form the basis of our subsequent perturbation analysis of heterogeneity and noise, we will outline the existence and stability analysis of bumps first, for an arbitrary number of layers $N$. The reader is advised to consult the works of Folias and Ermentrout for a more detailed characterization of the possible bifurcations of stationary patterns in a pair of neural fields~\cite{folias11,folias12}. We also note that, while we are restricting our analysis to the case of one-dimensional domains, we expect our results to extend to two or more dimensions as demonstrated in \cite{poll15}. Furthermore, previous experiments in rats have probed the behavior and neurophysiological underpinnings of spatial navigation along linear tracks~\cite{battaglia04,yoon16}. Thus, we believe the model we analyze here would be pertinent to these cases in which the environment is nearly one-dimensional. After we characterize the stability of stationary bump solutions, we will consider the effects of weak perturbations to these solutions, which will help reveal how noise, heterogeneity, and interlaminar coupling shape the network's processing of velocity inputs.

\subsection{Existence of bump solutions} 
\label{bumpexs}
In the absence of a velocity signal $(v(t) \equiv 0)$ and heterogeneity ($h_j(x) \equiv 0, \  \forall j$), we can characterize stationary solutions to Eq.~(\ref{nfmodel}), given by $u_j(x,t) = U_j(x)$. Conditions for the existence of stable stationary bumps in single layer neural fields have been well-characterized~\cite{amari77,laing02,guo05,bressloff12}, but much remains in terms of understanding how the form of $w_{jk}(x-y)$ would impact the existence and stability of bumps in a multilayer network. Furthermore, the stationary equations for bump solutions are a form of the well-studied Hammerstein equation~\cite{hammerstein30,anello04}, and bump stability is characterized by Fredholm integral equations of the second kind~\cite{atkinson76}. For our purposes, we will construct bumps under the assumption that they exist. Then, we will employ self-consistency, to determine solution validity. This is straightforward in the case of a Heaviside nonlinearity $f(u) = H(u - \theta)$, Eq.~(\ref{fheav}), but we can derive some results for general nonlinearities $f(u)$. First, note that, in the case of translationally symmetric kernels $w_{jk}(x-y)$, we obtain the following convolution relating stationary solutions $U_j(x)$ in each layer to one another:
\begin{equation}
\label{stateq}
U_j(x) = \sum_{k=1}^N \int_{- \pi}^{\pi} w_{jk}(x-y)f(U_k(y)) \d y, \hspace{5mm} j=1,...,N.
\end{equation}
In later analysis, we will also find the formula for the spatial derivative useful:
\begin{align}
U_j'(x) = \sum_{k=1}^N \int_{- \pi}^{\pi} \frac{\d}{\d x} w_{jk}(x-y)f(U_k(y)) \d y, \hspace{5mm} j=1,...,N.  \label{spatder}
\end{align} 
Next, since each $U_j(x)$ must be periodic in $x \in [- \pi, \pi]$, we can expand it in a Fourier series
\begin{align}
U_j(x) = \sum_{l = 0}^{M} A_{lj} \cos (lx) + \sum_{m =1}^{M} B_{mj} \sin (mx),  \label{Ujfour}
\end{align}
where $M$ is the maximal integer index of a mode for bumps in all $N$ layers. Indeed, there will be a finite number of terms in the Fourier series, Eq.~(\ref{Ujfour}), under the assumption that the weight functions $w_{jk}(x-y)$ all have a finite Fourier expansion. Since most typical smooth weight functions are well approximated by a few terms in a Fourier series~\cite{veltz10}, we take this assumption to be reasonable. Once we do so, we can construct solvable systems for the coefficients of the bumps, Eq.~(\ref{Ujfour}), and their stability as in \cite{carroll14}. For even symmetric weight kernels, we can write
\begin{align*}
w_{jk}(x-y) = \sum_{m=0}^M C_{jkm} \cos (m(x-y)) =  \sum_{m=0}^M C_{jkm} \left[ \cos (mx) \cos (my) + \sin (mx) \sin (my) \right],
\end{align*}
so that Eq.~(\ref{stateq}) implies that
\begin{subequations}  \label{ABcoeff}
\begin{align}
A_{lj} &= \sum_{k=1}^N C_{jkl} \int_{- \pi}^{\pi} \cos (lx) f(U_k(x)) \d x, \\
B_{mj} &= \sum_{k=1}^N C_{jkm} \int_{- \pi}^{\pi} \sin (mx) f(U_k(x)) \d x.
\end{align}
\end{subequations}
Since the noise-free, heterogeneity-free system is translationally invariant, there is a family of solutions with center of mass at any location on $x \in [ - \pi, \pi]$. Furthermore, the evenness of the weight functions $w_{jk}(x-y)$ we have chosen implies the resulting system is reflection symmetric, so we can restrict our examination to even solutions, so $B_{mj} \equiv 0$ for all $m,j$, so Eq.~(\ref{Ujfour}) becomes
\begin{align}
U_j(x) = \sum_{l=0}^{N} A_{lj} \cos (x).  \label{Ujcos}
\end{align}
Plugging the formula Eq.~(\ref{Ujcos}) into Eq.~(\ref{ABcoeff}), we find
\begin{align}
A_{lj} =  \sum_{k=1}^N C_{jkl} \int_{- \pi}^{\pi} \cos (lx) f \left( \sum_{m=0}^{N} A_{mk} \cos (mx) \right) \d x. \label{genAcoeff}
\end{align}
The coefficients $A_{lj}$ can be found using numerical root finders~\cite{veltz10}. However, for particular functions $f$ and $w_{jk}$, we can project the system Eq.~(\ref{genAcoeff}) to a much lower-dimensional set of equations, which can sometimes be solved analytically.

For instance, consider the Heaviside nonlinearity $f(u) = H(u - \theta)$, Eq.~(\ref{fheav}). In this case, stationary bump solutions $u_j(x,t) = U_j(x)$ centered at $x=0$ are assumed to have superthreshold activity on the interval $x \in [-a_j,a_j]$ in each layer $j=1,...,N$; i.e. $U_j(x) > \theta$ for $x \in [-a_j,a_j]$. Applying this assumption to the stationary Eq.~(\ref{stateq}) yields
\begin{align*}
U_j(x) = \sum_{k=1}^N \int_{- a_k}^{a_k} w_{jk}(x-y) \d y. 
\end{align*}
Self-consistency then requires that $U_j( \pm a_j) = \theta$, as originally pointed out by Amari~\cite{amari77}, which allows us to write
\begin{align}
\theta = \sum_{k=1}^N \int_{-a_k}^{a_k} w_{jk}(a_j-y) \d y,  \ \ \ \ j=1,...,N.  \label{bumpthresh}
\end{align}
Again, Eq.~(\ref{bumpthresh}) is a system of nonlinear equations, which can be solved numerically via root-finding algorithms. However, as opposed to the integral terms in Eq.~(\ref{genAcoeff}), the integrals in Eq.~(\ref{bumpthresh}) are tractable, which makes for a more straightforward implementation of a root-finder. If we utilize the canonical cosine weight functions, Eq.~(\ref{wcos}) and (\ref{winter}), we find we can carry out the integrals in Eq.~(\ref{bumpthresh}) to yield:
\begin{align}
\theta &= \sin (2 a_j) + \sum_{k \neq j} 2 \bar{w}_{jk} \left[ a_k + \cos (a_j) \sin (a_k)  \right].  \label{bwideq}
\end{align}
Henceforth, we mostly deal with the specific case of cosine weight connectivity, although we suspect our results extend to the case of more general weight functions. This allows us to define connectivity simply using the scalar strength values of the interlaminar coupling, which comprise the off-diagonal entries of the following matrix: ${\mc W}_{jk} = \left\{ \bar{w}_{jk}: j \neq k; \ \ 1: j=k\right\}$ for $j,k=1,...,N$. As discussed in Section \ref{sec:model}, and specifically Fig. \ref{fig1}B, we categorize the network graphs of primary interest to our work here into the main cases of a two-layer network and some specific cases of a network with more layers. We now briefly demonstrate how such graph structures can impact the stationary solutions, as it foreshadows the impact on the non-equilibrium dynamics of the network. 

\begin{figure}
\begin{center} \includegraphics[width=15.5cm]{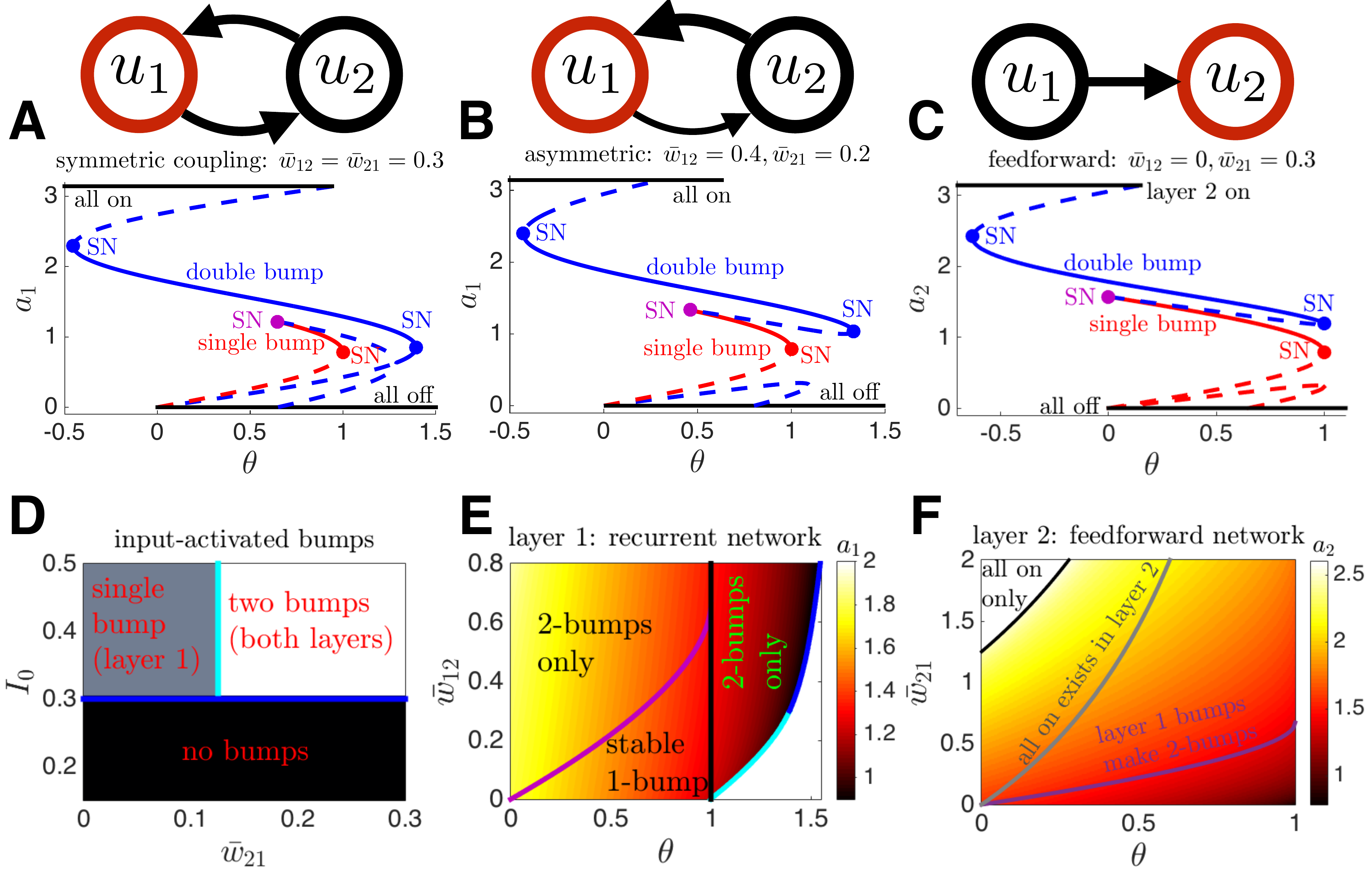} \end{center}
\caption{Bump half-width plots for two-layer ($N=2$) networks with Heaviside nonlinearity $f(u) = H(u - \theta)$, Eq.~(\ref{fheav}), and cosine coupling functions Eq.~(\ref{wcos}) and (\ref{winter}), as given by Eq.~(\ref{bwideq}). ({\bf A},{\bf B},{\bf C}) Bifurcation diagrams for half-width of bumps in the red layer shown in network diagram above. ({\bf A}) Half-width $a_1$ of the bump in layer $u_1$ in a symmetric network, plotted as a function of threshold $\theta$, as given by Eq.~(\ref{asymbwid}). Stable (solid) and unstable (dashed) branches of double bumps annihilate in a saddle-node (SN) bifurcation at low threshold $\theta$, and at high threshold $\theta$. Stable and unstable branches of single bumps also emerge from a SN for sufficiently high $\theta$, while the stable branch annihilates with a branch of double bumps for low enough $\theta$. ({\bf B}) Half-width $a_1$ as a function of $\theta$ in an asymmetric network, as given by Eq.~(\ref{asymbwid}). ({\bf C}) Bump half-width $a_2$ in a feedforward network, given by the single Eq.~(\ref{ffbwid}), shows both single bumps and double bump branches annihilate at the same upper threshold $\theta = 1$. ({\bf D}) Critical coupling $\bar{w}_{21}$ and input strength $I_0$ for $I(x) = I_0 \cos (x)$ needed to instantiate a single bump in layer 1 or a double bump solution, in a feedforward network, where $\theta = 0.3$. Shaded regions are generated by numerically simulating Eq.~(\ref{nfmodel}), and thick blue lines are calculated theoretically (see `Critical input needed for activation of bumps' in main text). ({\bf E}) Half-width $a_1$ of the layer 1 bump of a double bump solution for a recurrent network with $\bar{w}_{21}=0.3$ over a range of coupling strength $\bar{w}_{12}$ and threshold $\theta$. Partitions demonstrate that a stable 1-bump solution also coexists in a subregion of the domain. No 2-bumps exist in the white region. ({\bf F}) Half-width $a_2$ of the layer 2 bump of a double bump solution for a feedforward network over a range of coupling strength $\bar{w}_{21}$ and threshold $\theta$. Stable 1-bumps exist below the magenta line. For sufficiently large coupling $\bar{w}_{21}$ and low threshold, only the `all-on' solution exists in layer 2.}
\label{fig2}
\end{figure}

{\em Two-layer symmetric network ($\bar{w}_{12} \equiv \bar{w}_{21} = \bar{w}$).} In this case, we can derive a few analytical results concerning the bifurcation structure of stationary bump solutions. However, to identify the half-widths $a_1$ and $a_2$, it is typically necessary to solve Eq.~(\ref{bwideq}) numerically to produce the plots shown in Fig. \ref{fig2}A. First of all, for double bump solutions, in which both layers possess stationary superthreshold activity, if we assume symmetric solutions, so that $a_1 = a_2 = a$, then we can write Eq.~(\ref{bwideq}) as
\begin{align}
\theta &= ( 1 + \bar{w}) \sin (2 a) + 2 \bar{w} a \equiv G(a). \label{symbwid}
\end{align}
We cannot solve the transcendental Eq.~(\ref{symbwid}) explicitly for the bump half-width $a$. In order to gain some insight, we can identify the range over which solutions to the equations exist. This can be determined explicitly by finding the turning points of the right hand side of Eq.~(\ref{symbwid}) (See blue dots in Fig. \ref{fig2}A,B,C), corresponding to the extrema of the function between which solutions exist.  Thus, we can determine the location of these turning points, which are saddle-node (SN) bifurcations, by differentiating the right hand side $G(a)$:
\begin{align*}
G'(a) &= 2 (1 + \bar{w}) \cos (2a) + 2 \bar{w},
\end{align*}
so by requiring $G'(a_c) = 0$, we have
\begin{align*}
a_c = \frac{1}{2} \cos^{-1} \left[ \frac{\bar{w}}{1+ \bar{w}} \right], \ \pi - \frac{1}{2} \cos^{-1} \left[ \frac{\bar{w}}{1+ \bar{w}} \right],
\end{align*}
matching the locations of the double bump SN bifurcations (blue dots) shown in Fig. \ref{fig2}A.

Furthermore, SN bifurcations associated with the coalescing of stable single bump branches with unstable double bump branches (purple dots in Fig. \ref{fig2}A,B,C) can be determined using a threshold condition. For instance, given a layer 1 bump with half-width $a_1$, we require the stationary solution in layer 2 ($u_2 = U_2(x)$) remains subthreshold ($U_2(x) < \theta$, $x \in [- \pi,\pi]$). Given $a_2 =0$ in Eq.~(\ref{bwideq}), single bump solutions in layer 1 satisfy $\theta = \sin (2a_1)$, so $a_u = \frac{1}{2} \sin^{-1} \theta, a_s =\frac{\pi}{2} - \frac{1}{2} \sin^{-1} \theta$ are solutions with $a_s$ corresponding to the stable bump~\cite{kilpatrick13}. Thus, we require $U_2(x) = 2 \bar{w} (a_1 + \sin (a_1) \cos (x))< \theta$, so selecting for the maximal value of $U_2(x)$ and plugging in $a_s$, we have an explicit equation for the critical interlaminar strength $\bar{w}$ above which there are no stable single bump solutions: $\bar{w}_c  = \sin(2a_s)/ \left[ a_s + \sin (a_s) \right]$, providing an implicit equation for the SN locations in Fig. \ref{fig2}A,B,C, and corresponding to the magenta curves in Fig. \ref{fig2}E,F.

{\em Two-layer asymmetric network ($\bar{w}_{12} \neq \bar{w}_{21}$).} Double bump solution half-widths tend to differ in this case $a_1 \neq a_2$, obeying the pair of implicit equations
\begin{subequations}  \label{asymbwid}
\begin{align}
\theta &= \sin (2 a_1) + 2 \bar{w}_{12} \left[ a_2 + \cos (a_1) \sin (a_2)  \right], \\
\theta &= \sin (2 a_2) + 2 \bar{w}_{21} \left[ a_1 + \cos (a_2) \sin (a_1)  \right],
\end{align}
\end{subequations}
which we solve numerically to generate the branches plotted in Fig. \ref{fig2}B, as well as the surface plot in Fig. \ref{fig2}E. Note, however, it is still possible to determine the range of values in which stable single bump solutions exist in layer $j$ using the requirement $\bar{w}_{kj} < \sin(2a_s)/ \left[ a_s + \sin (a_s) \right]$, as derived in the symmetric network case.

{\em Two-layer feedforward network ($\bar{w}_{12} \equiv 0$).} This is a special case of the asymmetric network, where the nonlinear system, Eq.~(\ref{bwideq}), defining the bump half-widths reduces to:
\begin{align*}
\theta = \sin (2a_1),  \hspace{5mm} \theta = \sin (2a_2) + 2 \bar{w}_{21} \left[ a_1 + \cos (a_2) \sin (a_1) \right],
\end{align*}
which can further be reduced to a single implicit equation for the half-width in the target layer 2 (see schematic in Fig. \ref{fig2}C):
\begin{align}
\theta = \sin (2a_2) + 2 \bar{w}_{21} \left[ \frac{\pi}{2} - \frac{1}{2} \sin^{-1} \theta + \frac{\cos (a_2)}{2} \left( \sqrt{1-\theta} + \sqrt{1+ \theta} \right) \right],  \label{ffbwid}
\end{align}
which can be solved using numerical root finding to yield the curves in Fig. \ref{fig2}C,F.

{\em `All-on' solutions in the two-layer network.} Given excitatory interlaminar connections, it is possible to generate `all-on' solutions in one and sometimes two layers of the network. An `all-on' solution is one in which a layer has a stationary solution $U_j(x)$ that is entirely superthreshold, $U_j(x) > \theta$ for all $x \in [- \pi, \pi]$. In the case of a feedforward network (Fig. \ref{fig2}C,F), the target layer 2 will have an `all-on' solution when the minimal value of $U_2(x)>0$ given a stable bump solution $U_1(x)$ in layer 1. As a result, an `all-on' solution in layer 2 would have the form
\begin{align*}
U_2(x) = 2 \bar{w}_{21} \left[ a_1 + \sin (a_1) \cos (x) \right],
\end{align*}
so requiring ${\rm min}_x[U_2(x)]>\theta$ yields
\begin{align*}
\bar{w}_{21} \left[ \pi  -  \sin^{-1} \theta - \sqrt{1- \theta} - \sqrt{1+\theta} \right]> \theta,
\end{align*}
obtaining equality along the grey line plotted in Fig. \ref{fig2}F. For recurrent networks, we can easily identify the threshold curves $(\bar{w}_{jk},\theta)$ above which double `all-on' solutions exist. These have the simpler forms:
\begin{align*}
U_1(x) = 2 \bar{w}_{12} \pi, \hspace{5mm} U_2(x) = 2 \bar{w}_{21} \pi,
\end{align*}
so we need to require that $\bar{w}_{12} > \theta/(2 \pi)$ and $\bar{w}_{21} > \theta/(2 \pi)$.

{\em Critical input needed for activation of bumps.} We are studying multilayer networks wherein we assume bump solutions can be instantiated by an external input. However, it is important to identify the critical input needed to nucleate and maintain such bumps in the two layers of the network. As demonstrated in Fig. \ref{fig2}A,B,C, there are multiple stable stationary solutions across a range of threshold $\theta$ and coupling values $(\bar{w}_{12}, \bar{w}_{21})$.

We wish to demonstrate that it is possible to instantiate a two bump solution given only an input, $I(x) = I_0 \cos (x)$, to layer 1, and we focus exclusively on the feedforward network. This single layer will only have subthreshold activity if $I_0 \cos (x) < \theta$ everywhere. If input is superthreshold ($I_0> \theta$), stationary bump solutions, driven by an input in layer 1, are then given~\cite{hansel98,folias04}: $U_1(x) = \left[ 2 \sin (a_1) + I_0 \right] \cos (x)$. Thus, bumps driven by inputs just beyond the critical level $I_0 = \theta$, will have half-widths approximately satisfying $\theta = \sin (2a_1) + \theta \cos (a_1)$. These bumps will have half-widths then given by the implicit equation $\theta (a_1) := \sin (2a_1)/(1- \cos (a_1)) = 2 \cos (a_1) \cot (a_1/2)$. For values of $\theta (a_1)>0$ with $a_1 \in [0,\pi]$, we can show that this function is monotone decreasing, since $\theta'(a_1) = - 2 \cos (a_1) - \csc^2(a_1/2) <0$ when $0< a_1< a_1^c \approx 2.2372$. In this case, $\theta (a_1) \approx -0.6006$. Therefore, as $\theta (a_1)$ will tend to increase as $a_1$ is decreased from $a_1^c$, so for $\theta >0$ we expect a single stable stationary bump solution in layer 1 (See also \cite{folias04,kilpatrick13}). This suggests either single or double bump solutions will emerge as long as $I_0> \theta$, as shown in Fig. \ref{fig2}D. Increasing the strength of input $I_0$ will only serve to further stabilize this stationary bump. To determine the critical strength needed to propagate this bump forward to layer 2, we must solve for the half-width $a_1$ in $\theta = \sin (a_1) + I_0 \cos (a_1)$, and require that layer 2 is driven superthreshold, so that $U_2(0) = 2 \bar{w}_{21} \left[ a_1 + \sin (a_1) \right] > \theta$. This admits an explicit inequality $\bar{w}_{21} > \frac{\theta}{2 \left[ a_1 + \sin (a_1)\right]}$, so we need only solve for $a_1$ numerically to obtain the vertical boundary between single and double bump solutions in Fig. \ref{fig2}D.

\subsection{Linear stability of bumps}
\label{blinstab}
Linear stability of the bump solutions $U_j(x)$, Eq.~(\ref{stateq}), can be determined by analyzing the evolution of small, smooth, separable perturbations such that $u_j(x,t) = U_j(x) + \ve \e^{\lambda t} \psi_j (x)$. We expect $u_j = U_j(x)$ ($j=1,...,N$) to be neutrally stable to translating perturbations $\psi_j = U_j'(x)$, arising from the translation symmetry of Eq.~(\ref{nfmodel}) given $w_{jk}(x,y) = w_{jk}(x-y)$. On the other hand, the bump may be linearly stable or unstable to perturbations of its half-width $a_j$~\cite{amari77}. The results we derive here for such perturbations are what determine the stability of branches plotted in Fig. \ref{fig2}.

To begin, consider $u_j(x,t) = U_j(x) + \ve \Psi_j(x,t)$, where $\Psi_j(x,t)$ 
thus describes perturbations to the shape of the bump $U_j(x)$ that may evolve temporally. Plugging this into the full neural field Eq.~(\ref{nfmodel}) with $h_j \equiv 0$, $Z_j \equiv 0$, and $v\equiv 0$, we can apply the stationary Eq.~(\ref{stateq}), and subsequently write the ${\mc O}(\ve)$ equation as
\begin{align}
\frac{\pd \Psi_j(x,t)}{\pd t} &= - \Psi_j(x,t) + \sum_{k=1}^N \int_{- \pi}^{\pi} w_{jk}(x-y) f'(U_k(y)) \Psi_k(y,t) \d y.   \label{psi1}
\end{align}
Due to the linearity of the equation, we may apply separation of variables to each $\Psi_j$, such that $\Psi_j(x,t) = b_j(t) \psi_j(x)$~\cite{sandstede02,veltz10}. Substituting into Eq.~(\ref{psi1}), we have for each $j=1,...,N$:
\begin{align}
\dfrac{b_j'(t)}{b_j(t)} = -1 + \dfrac{1}{\psi_j(x)} \sum_{k=1}^N \int_{- \pi}^{\pi} w_{jk}(x-y)f'(U_k(y)) \psi_k(y) \d y.  \label{psi2}
\end{align}
Thus, each side of Eq.~(\ref{psi2}) depends exclusively on a different variable, $x$ or $t$, so both must equal a constant $\lambda$. Therefore, $b_j(t) = c_j \e^{\lambda t}$ for all $j=1,...,N$, suggesting perturbations will grow indefinitely as $t \to \infty$ for ${\rm Re} \lambda >0$, indicating instability. While oscillatory instabilities are plausible (${\rm Re}\lambda >0$ with ${\rm Im} \lambda \neq 0$), given specific forms of interlaminar coupling (e.g., combinations of interlaminar excitation and inhibition~\cite{folias12}), we did not identify such instabilities in the mutual excitatory layered networks we studied (Fig. \ref{fig2}). Thus, we expect instabilities emerging where ${\rm Re} \lambda = 0$ will typically be of saddle-node type (${\rm Im} \lambda =0$). Furthermore, the equation for $\psi_j(x)$ is now given for all $j=1,...,N$, as
\begin{align}
(\lambda + 1) \psi_j(x) = \sum_{k=1}^N \int_{- \pi}^{\pi} w_{jk}(x-y)f'(U_k(y))\psi_k(y) \d y. \label{psi3}
\end{align}
Eigenvalues $\lambda$ are thus determined by consistent solutions $(\lambda, \bpsi (x))$ for $\bpsi = ( \psi_1, \psi_2,...,\psi_N)^T$, to Eq.~(\ref{psi3}). One such solution is $(\lambda, \bpsi (x)) = (0,\mathbf{U}'(x))$ for $\mathbf{U}' = (U_1',U_2',...,U_N')^T$, as can be shown by applying Eq.~(\ref{spatder}). As mentioned above, this demonstrates the neutral stability of bump solutions to translating perturbations, due to the translational invariance of Eq.~(\ref{nfmodel}).

Further analysis in the case of a general firing rate function $f(u)$ can be difficult. However, if we consider the Heaviside nonlinearity $f(u) = H(u - \theta)$ given by Eq.~(\ref{fheav}), we obtain a specific case of Eq.~(\ref{psi3}), which is easier to analyze~\cite{amari77,coombes04}:
\begin{align}
(\lambda + 1) \psi_j(x) = \sum_{k=1}^N \gamma_k \bigg( w_{jk}(x - a_k) \psi_k(a_k) + w_{jk}(x + a_k) \psi_k(-a_k) \bigg),  \label{psi4}
\end{align}
where we have made use of the fact that
\begin{align}
f'(U_j(x)) = \delta (U_j(x) - \theta) = \frac{\delta (x+a_j)}{U_j'(-a_j)} -  \frac{\delta (x-a_j)}{U_j'(a_j)} =  \frac{\delta (x+a_j) + \delta (x-a_j)}{|U_j'(a_j)|},  \label{fprime}
\end{align}
since $U_j'(-a_j) = -U_j(a_j)>0$, and we have assigned
\begin{align}
\gamma_j^{-1} = |U'(a_j)| = \sum_{k=1}^N \left[ w_{jk}(a_j - a_k) - w_{jk}(a_j + a_k) \right], \hspace{5mm} j=1,...,N, \label{gammak}
\end{align}
under the assumption that $w_{jk}(x)$ is monotone decreasing in $|x|$, which is the case for cosine weight functions, Eq.~(\ref{wcos}) and (\ref{winter}). Note, all eigenfunctions $\bpsi(x)$ of Eq.~(\ref{psi4}) that satisfy the conditions $\psi_j(\pm a_j) =0$, for all $j=1,...,N$, have associated eigenvalue given by $(\lambda + 1) \bpsi = 0$ so $\lambda = -1$, which does not contribute to any instabilities. To specify other eigensolutions, we examine cases where $\psi_j(\pm a_j)  \neq 0$ for at least one $j=1,...,N$. In such cases, we can obtain expressions for the eigenvalues by examining Eq.~(\ref{psi4}) at the points $x=\pm a_1, \pm a_2, ..., \pm a_N$. In this case, the eigenfunctions $\bpsi$ are defined by their values at the threshold crossing points: $\psi_j(\pm a_j)$ for each $j=1,...,N$. Thus, defining these unknown values $A_j^{\pm} : = \psi_j(\pm a_j)$, we simplify Eq.~(\ref{psi4}) to a linear system of $2N$ equations of the form 
\begin{align}
(\lambda + 1) \v = \W \v, \hspace{4mm} \v = (A_1^+,...,A_N^+, A_1^-, ...,A_N^-)^T, \hspace{4mm} \W = \begin{bmatrix} {\mc A}_- &{\mc A}_+ \\ {\mc A}_+ &{\mc A}_- \end{bmatrix},  \label{psieval}
\end{align}
where the elements of the blocks of the $2N \times 2N$ matrix $\W$ are $\left( \mathcal{A}_{\pm} \right)_{jk} = \gamma_k w_{jk}(a_j \pm a_k)$. We make use of the result $|\W|=|{\mc A}_- + {\mc A}_+||{\mc A}_- - {\mc A}_+|$~\cite{silvester00}, which implies the set of eigenvalues $\lambda_W$ of $\W$ is the union of the set of the eigenvalues of ${\mc A}_- + {\mc A}_+$ and ${\mc A}_- - {\mc A}_+$. Subsequently, the eigenvalues of $\W - I$ will be $\lambda = \lambda_W -1$. We now outline a few examples in which we can compute these eigenvalues analytically.

\begin{figure}
\begin{center} \includegraphics[width=12cm]{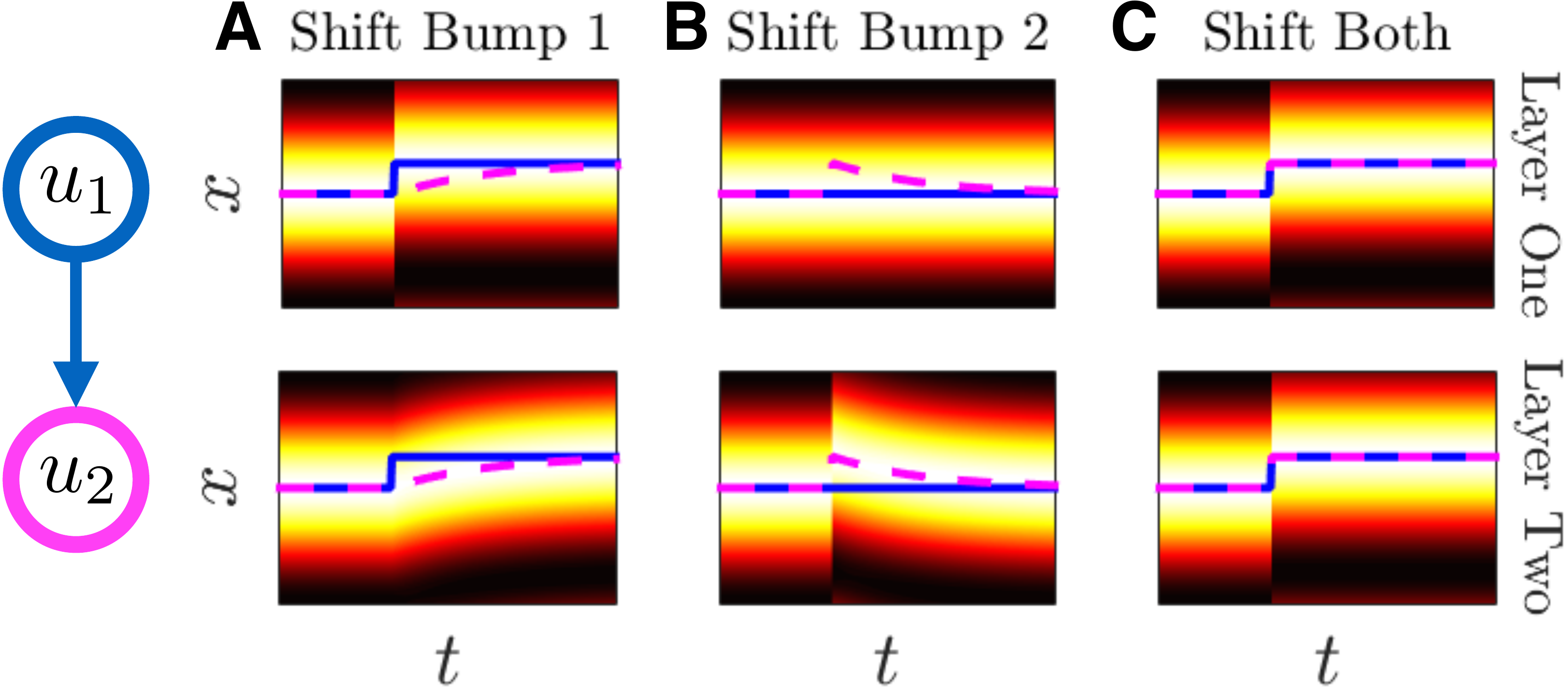} \end{center}
\caption{Linear stability of bumps in a feedforward two-layer network, demonstrated by simulations of the model Eq.~(\ref{nfmodel}) with $N=2$. ({\bf A}) When the bump in layer 1 is shifted, the bump in layer 2 (dashed line) relaxes to the new position of bump 1 (solid line). ({\bf B}) When the bump in layer 2 is shifted, it relaxes back to the fixed position of bump 1. ({\bf C}) When both bumps are shifted, both retain their new position, respecting the translation symmetry of the underlying Eq.~(\ref{nfmodel}).}
\label{fig3}
\end{figure}

{\em Two-layer feedforward network.} Assuming $w_{12} \equiv 0$, layer 1 sends input to layer 2, but receives no feedback from layer two. Linear stability associated with the stationary bump solutions to this model is then determined in part by computing the eigenvalues of:
\begin{align}
{\mc A}_- + {\mc A}_+ = \begin{bmatrix}
\gamma_1 w_{11}^+ &0 \\
\gamma_1 w_{21}^+ &\gamma_2 w_{22}^+
\end{bmatrix}, \; \; \; {\mc A}_- - {\mc A}_+ = \begin{bmatrix}
\gamma_1 w_{11}^- &0 \\
\gamma_1 w_{21}^- &\gamma_2 w_{22}^-
\end{bmatrix},  \label{ffmats}
\end{align}
where $w_{jk}^\pm := \gamma_j \big( w_{jk}(a_j - a_k) \pm w_{jk}(a_j + a_k) \big)$. Since the matrices in Eq.~(\ref{ffmats}) are triangular, their eigenvalues $\lambda_W$ are given by their diagonal entries. Applying Eq.~(\ref{gammak}), $\gamma_j^{-1} = \sum_{k=1}^N w_{jk}^-$, we can express eigenvalues $\lambda = \lambda_W  - 1$ of $\W - I$ as:
\begin{align*}
\lambda = \{ \lambda_1^-, \lambda_2^-, \lambda_1^+, \lambda_2^+ \} =  \left\{ 0, -\dfrac{w_{21}^-}{w_{21}^- + w_{22}^-} , \dfrac{2w_{11}(2a_1)}{w_{11}^-}, \dfrac{2w_{22}(2a_2) - w_{21}^-}{w_{21}^- + w_{22}^-} \right\}.
\end{align*}
Neutral stability with respect to the eigenfunction $\bpsi = {\mathbf U}'$ ensures the existence of $\lambda_1^- = 0$. Concerning the other three eigenvalue formulae, the terms $w_{jk}^-$ will be positive by our assumptions on the weights $w_{jk}$ made after Eq.~(\ref{gammak}), so $\lambda_2^- <0$, corresponding to the fact that the bump in layer 2 is linearly stable to translating perturbations, when the position of the bump in layer 1 is held fixed. In Fig. \ref{fig3}, we show that the upstream layer (1) governs the long term location of both bumps. The layer 2 bump always relaxes to the layer 1 bump's location. In a related way, the eigenvalues $\lambda_1^+$ and $\lambda_2^+$ correspond to expansions/contractions of the bump widths in layers 1 and 2, respectively. Typically, there are two bump solutions in a single-layer network, whose width perturbations correspond with the eigenvalue $\lambda_1^+$: one that is narrow and unstable to such perturbations ($\lambda_1^+>0$), and another that is wide and stable to such perturbations ($\lambda_1^+<0$)~\cite{amari77,coombes04,kilpatrick13}. Lastly, the bump in layer 2, driven by activity in layer 1 is influenced by features of layers 1 and 2, as shown in the formula for $\lambda_2^+$. When $2 w_{22}(2a_2)<0$, we expect $\lambda_2^+<0$, and the bump will be stable to width perturbations.

{\em Exploding star network.} Another example architecture involves a single layer with feedforward projections to multiple ($N-1$) layers. In this case, $w_{jk} \equiv 0$ for $j=2,...,N$ and $k \neq j$. Only perturbations that shift the bump in layer 1 have a long term impact on the position of bumps in the network. The translation modes of bumps in layers $j=2,...,N$ have associated negative eigenvalues, as we will show, which is a generalization of the two-layer feedforward case. Linear stability is computed by first determining the eigenvalues of:
\begin{align}
{\mc A}_- \pm {\mc A}_+ = \begin{bmatrix}
\gamma_1 w_{11}^{\pm} &0 &... &0 \\
\gamma_1 w_{21}^{\pm} & \gamma_2 w_{22}^{\pm} &... &0 \\
\vdots &\vdots &\ddots &\vdots \\
\gamma_1 w_{N1}^{\pm} &0 &... &\gamma_N w_{NN}^{\pm}
\end{bmatrix}.  \label{expstareig}
\end{align}
Subtracting one from the eigenvalues of the matrices in Eq.~(\ref{expstareig}) and applying the formula for $\gamma_j$, Eq.~(\ref{gammak}), we find $2N$ eigenvalues, given by $\lambda_j^{\pm}$ for $j=1,...,N$, where
\begin{align*}
\lambda_1^- = 0, \hspace{3mm} \lambda_1^+ = \frac{2w_{11}(2a_1)}{w_{11}^-}, \hspace{3mm} \lambda_j^{-} = - \frac{w_{j1}^-}{w_{j1}^-+w_{jj}^-}, \hspace{3mm} \lambda_j^+ = \frac{2w_{jj}(2a_j) - w_{j1}^-}{w_{j1}^- + w_{jj}^-}, \hspace{2mm} j=2,...,N.
\end{align*}
As in the two-layer network, bumps are neutrally stable to perturbations of the form $\bpsi = {\bf U}'$, corresponding to $\lambda_1^- = 0$. In addition, we expect $\lambda_j^-< 0$ for $j=2,...,N$ since $w_{jk}^->0$. As mentioned above, we would expect wide bumps to be stable to expansion/contraction perturbations, whose stability is described by the eigenvalues $\lambda_j^+$ for $j=1,...,N$.

\begin{figure}
\begin{center} \includegraphics[width=12cm]{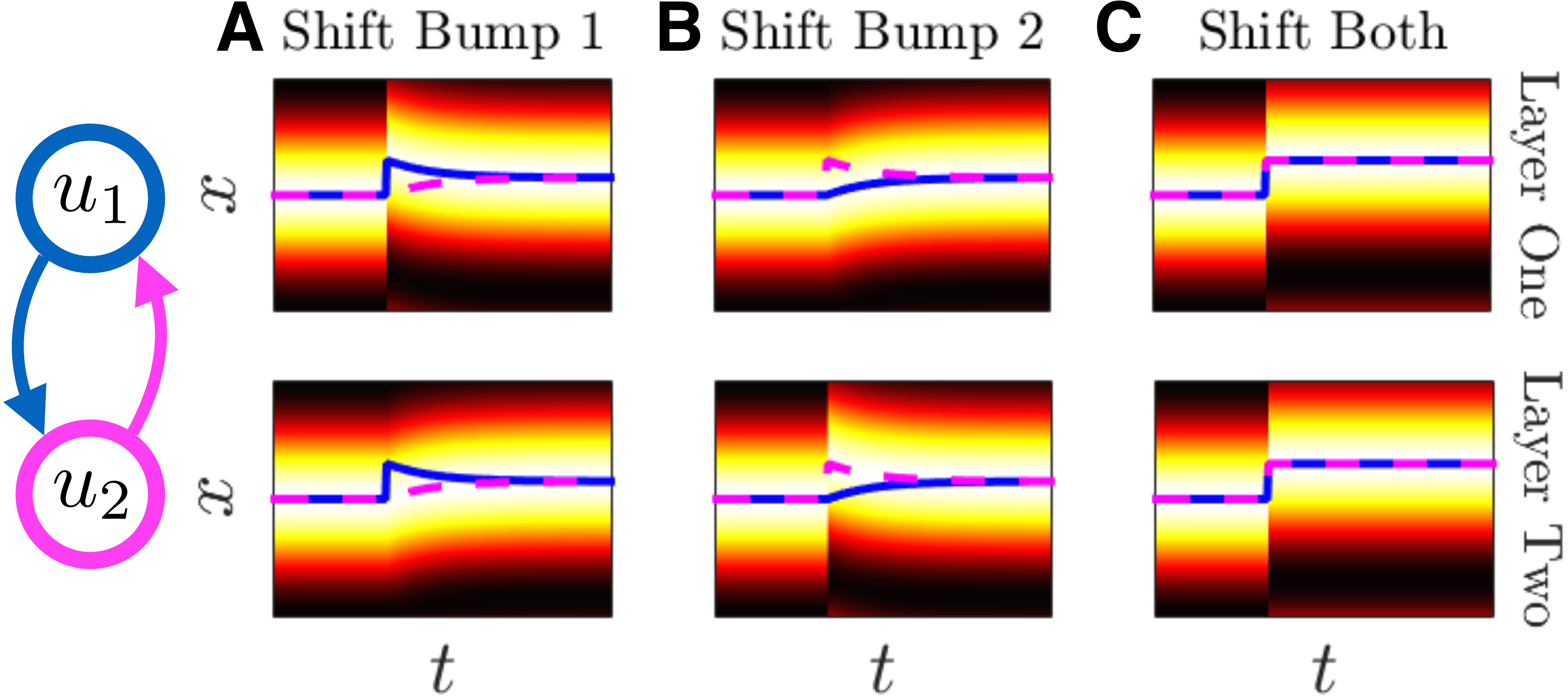} \end{center}
\caption{Linear stability of bumps in a two-layer symmetric recurrent network with $\bar{w}_{12} = \bar{w}_{21} >0$, demonstrated by simulations of the model Eq.~(\ref{nfmodel}) with $N=2$. Bumps initially at the same position $x=0$ are perturbed to examine the resulting evolution of their positions. ({\bf A}) When the bump in layer 1 (solid line) is shifted, both bumps relax to the average of their initially perturbed position. ({\bf B}) When the bump in layer 2 (dashed line) is shifted, again, both bumps relax to an intermediate position. ({\bf C}) When both bumps are shifted to the same location, both retain their new position.}
\label{fig4}
\end{figure}

{\em Two-layer recurrent network.} In the case of a fully recurrent network, where $\bar{w}_{jk}>0$ for all $j \neq k$, all matrix entries are nonzero: $\displaystyle {\mc A}_- \pm {\mc A}_+ = \begin{bmatrix}
\gamma_1 w_{11}^{\pm} &\gamma_2 w_{12}^{\pm} \\
\gamma_1 w_{21}^{\pm} &\gamma_2 w_{22}^{\pm}
\end{bmatrix}$. First, note that $\lambda_W=1$ is an eigenvalue of ${\mc A}_- - {\mc A}_+$, since $(w_{11}^-+w_{12}^-)(w_{22}^-+w_{21}^-) \cdot |{\mc A}_- - {\mc A}_+-I| = w_{12}^- w_{21}^- - w_{12}^-w_{21}^- = 0$, so we denote $\lambda_1^- = 0$ as the eigenvalue describing the translation symmetry of bumps. To gain further insight, we can also compute the other three eigenvalues:
\begin{align*}
\lambda_2^- &= - \dfrac{w_{12}^-}{w_{11}^-+w_{12}^-} - \dfrac{w_{21}^-}{w_{22}^-+w_{21}^-}, \\
\lambda_{1,2}^+ &=  \dfrac{ \gamma_1 w_{11}^+ + \gamma_2 w_{22}^+ -2 \pm \sqrt{\left[ \gamma_1 w_{11}^+ + \gamma_2 w_{22}^+ -2 \right]^2 - 4 \gamma_1 \gamma_2 \left[ (w_{11}^+ - \gamma_1^{-1}) (w_{22}^+ - \gamma_2^{-1}) - w_{12}^+w_{21}^+ \right] } }{2}.
\end{align*}
For a symmetric recurrent network: $w_{jj} \equiv w$, $w_{jk} \equiv w_c$, and $\gamma_j = \gamma$ ($j=1,2$, $k \neq j$), these formulas reduce to $\lambda_2^- = - \dfrac{2w_{c}^-}{w^-+w_c^-}$ and $\lambda^+ : = \lambda_{1,2}^+  =  \gamma w^+  -1 \pm w_{c}^+$. Bumps are linearly stable to perturbations that move them apart (Fig. \ref{fig4}A,B), and neutrally stable to translations that move them to the same location (Fig. \ref{fig4}C).

\begin{figure}
\begin{center} \includegraphics[width=12cm]{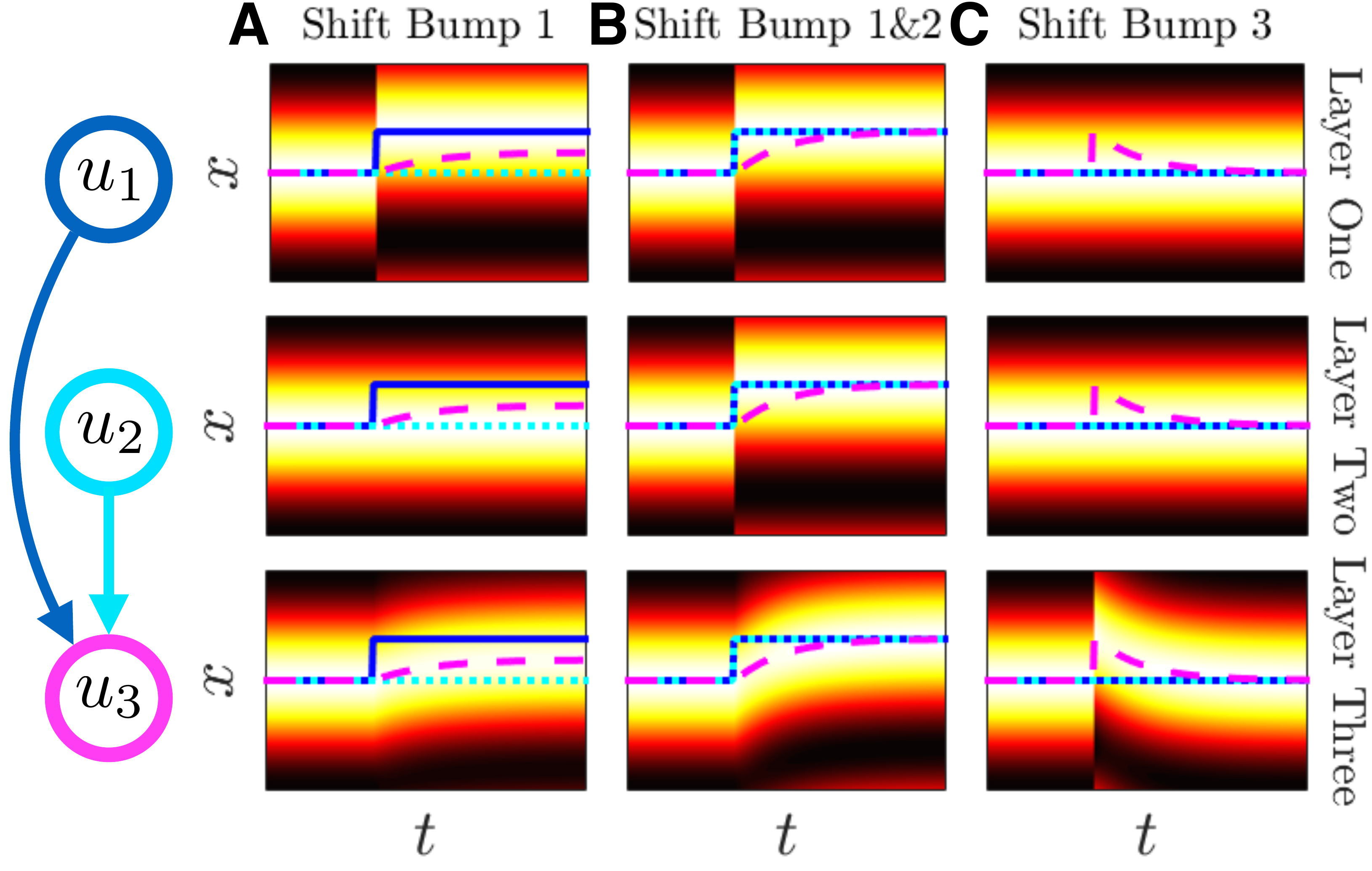} \end{center}
\caption{Linear stability of bumps in a $N=3$ layer imploding star network with $\bar{w}_{jk} \equiv 0$ for $j \neq 3$, $k \neq j$. ({\bf A}) When the bump in layer 1 (solid line) is shifted, the bump in layer 3 (dashed line) relaxes to a position between the layer 1 and layer 2 (dotted line) bump. ({\bf B}) When both bumps in layers 1 and 2 are perturbed to a new location, the bump in layer 3 relaxes to that new location. ({\bf C}) When only the bump in layer 3 is perturbed, it relaxes back to the locations of the bumps in layer 1 and 2.}
\label{fig5}
\end{figure}

{\em Imploding star graphs.} Additional dimensions of neutral stability can arise in the case of more than two layers, depending on the graph defining interlaminar connectivity. For instance, if there are multiple layers $j=1,...,l$ that receive no feedback from other layers, then $\gamma_j w_{jj}^- = 1$ for $j=1,...,l$, so the first $l$ rows of the matrix ${\mc A}_- - {\mc A}_+$ are the canonical unit vectors ${\bf e}_1,...,{\bf e}_l$. Thus, there are at least $l$ unity eigenvalues of $\W$, implying $\lambda = 0$ has multiplicity at least $l$ in Eq.~(\ref{psieval}), corresponding to the neutral stability of bumps in the $l$ layers that receive no feedback. We consider such an example when $N=3$:
\begin{equation*}
{\mc A}_- + {\mc A}_+ = \begin{bmatrix}
\gamma_1 w_{11}^+ &0 &0 \\
0 &\gamma_2 w_{22}^+ &0 \\
\gamma_1 w_{31}^+ &\gamma_2 w_{32}^+ &\gamma_3 w_{33}^+ \\
\end{bmatrix}, \; \; \; {\mc A}_- - {\mc A}_+ = \begin{bmatrix}
1 &0 &0 \\
0 & 1 &0 \\
\gamma_1 w_{31}^- &\gamma_2 w_{32}^-  &\gamma_3 w_{33}^-
\end{bmatrix}.
\end{equation*}
Eigenvalues of Eq.~(\ref{psieval}) are then $\lambda_{j}^- = 0$ and $\lambda_j^+ = \dfrac{ w_{jj}^+ - w_{jj}^-}{w_{jj}^-}$ for $j=1,2$, and
\begin{equation*}
\lambda_3^- = \dfrac{ - (w_{12}^- + w_{13}^-)}{ w_{11}^- + w_{12}^- + w_{13}^-}, \hspace{8mm} \lambda_3^+ = \dfrac{ w_{33}^+ - (w_{31}^- + w_{32}^- + w_{33}^-)}{w_{31}^- + w_{32}^- + w_{33}^-} .
\end{equation*}
Bumps in both layers 1 and 2 are neutrally stable to translations (Fig. \ref{fig5}A,B), whereas the bump in layer 3 is linearly stable to translation, relaxing to a weighted average of the positions of the layers 1 and 2 bumps (Fig. \ref{fig5}C). Adding dimensions to the space of translation symmetric perturbations will change the low-dimensional approximation that captures the dynamics of multilayer bump solutions in response to noise perturbations (Compare Sections \ref{dereff} and \ref{sec:ndb}).

{\em Directed loop of $N$ layers.} As a last example, we consider an $N$-layer directed loop, wherein each layer provides feedforward synaptic input to a subsequent layer. As a result, there is a band of nonzero interlaminar coupling along $w_{j+1,j}$ for $j=1,..,N$ (replace $N+1$ with $1$). Again, there is a zero eigenvalue $\lambda$ in Eq.~(\ref{psieval}), since
\begin{align*}
{\mc A}_- - {\mc A}_+ = \left[ \begin{array}{cccc}
\gamma_1 w_{11}^- &0 & \cdots & \gamma_N w_{1N}^- \\
\gamma_1 w_{21}^- &\gamma_2 w_{22}^- &0 & \cdots \\
0 & \ddots & \ddots & 0 \\
0 & 0 &\gamma_{N-1} w_{N,N-1}^- &\gamma_N w_{NN}^- \end{array} \right].
\end{align*}
Our desired result can be demonstrated by computing the determinant of the bidiagonal matrix:
\begin{align*}
|{\mc A}_- - {\mc A}_+ - I|  = \prod_{j=1}^N (\gamma_j w_{jj}^- -1) - \prod_{j=1}^N \left[ - \gamma_j w_{j+1,j} \right] = \prod_{j=1}^N \left[ - \gamma_j w_{j,j-1}^- \right] - \prod_{j=1}^N \left[ - \gamma_j w_{j+1,j}  \right] = 0,
\end{align*}
replacing $j-1=0$ with $N$ in the case $j=1$. We have applied the fact that $\gamma_j^{-1} = w_{jj}^- + w_{j,j-1}^-$ to transform the first product to the form of the second.

\subsection{Derivation of the effective equations}
\label{dereff}
Our stability analysis has provided us insight into the qualitative behavior of the multilayer bump solutions when small perturbations are applied. The underlying architecture both within and between layers shapes the response. We now extend our linear stability results to study the impact of persistent noise perturbations to stationary bump solutions, with heterogeneity as described by Eq.~(\ref{rechetero}) and velocity input described by Eq.~(\ref{velwt}). We begin by assuming that first, we only need to consider a single stochastically-evolving position, $\Delta(t)$, corresponding to the relative location of the entire multilayer bump solution. This assumes a single dimension of translation symmetry in the linear stability problem of the bump solution, computed in Section \ref{blinstab}. Cases in which more than one such dimension exists will be analyzed in Section \ref{sec:ndb}. Secondly, we assume a separation of timescales between the position and width perturbations of each bump, leading to the ansatz: $u_j(x,t) := U_j(x - \Delta (t)) + \ve \Phi_j(x-\Delta (t),t)$, where $\Phi_j$ describes the dynamics of shape perturbations to the bump in layer $j$. In line with previous studies of the impact of noise on patterns in neural fields~\cite{bressloff12,kilpatrick13}, the displacement of the bump from its initial position is assumed to be weak and slow, so that $\Delta (t)$ and $\d \Delta (t)$ are ${\mc O}(\ve)$. We find that the results of our perturbation analysis are consistent with this assumption. Since the spatial heterogeneity, velocity, and noise are all scaled by $\ve$, we expect them to enter into the derived effective equation. Note, in the case of weak interlaminar coupling, we would consider a separate stochastic variable $\Delta_j$ for each layer's bump~\cite{kilpatrick13b,bressloff15}. Plugging our ansatz into Eq.~(\ref{nfmodel}) and disregarding higher order terms $\mathcal{O}(\ve^2)$, the following equation in $\mathcal{O}(\ve)$ remains:
\begin{align}
\label{Leq}
\d \Phi_j(x,t) =& \bigg[  \mathcal{L}_j\big[ \bPhi(x,t)\big]  + \int_{- \pi}^{\pi} h_j(y+\Delta)w_{jj}(x-y)f(U_j(y)) \d y  \nonumber \\ 
& \ \left. +\; v(t) \sum_{k=1}^N \int_{- \pi}^{\pi} w_{vjk}(x-y)f(U_k(y)) \d y \right] \d t + \ve^{-1} \d \Delta U'_j(x)+  \d Z_j(x,t),
\end{align}
where $\mathcal{L}_j$ is the $j^{th}$ element of the linear functional $\mathcal{L}: \mathbf{p} \mapsto \mathbf{q}$ for $\mathbf{p}=(p_1,p_2,...,p_N)^T$ and $\mathbf{q}=(q_1,q_2,...,q_N)^T$, defined as
\begin{align*}
\mathcal{L}_j\big[ \mathbf{p}(x) \big] = -p_j(x) + \sum_{k=1}^N \int_{- \pi}^{\pi} w_{jk}(x-y)f'(U_k(y))p_k(y) \d y, \hspace{9mm} j=1,...,N,
\end{align*}
with adjoint operator $\mathcal{L}^*: \mathbf{q} \mapsto \mathbf{p}$, defined $\langle {\mc L} \mathbf{p}, \mathbf{q} \rangle = \langle \mathbf{p}, {\mc L}^*\mathbf{q} \rangle $ under the standard $L^2$ inner product, and thus given element-wise by
\begin{equation*}
\mathcal{L}^*_j\big[ \mathbf{q}(x)\big] = -q_j(x) + f'(U_j(x)) \sum_{k=1}^N \int_{- \pi}^{\pi} w_{kj}(x-y) q_k(y) \d y,  \hspace{9mm} j=1,...,N,
\end{equation*}
note the exchange in the order of the indices in $w_{kj}(x)$. Note also that Eq.~(\ref{Leq}) suggests that $\Delta$ and $\d \Delta$ should be ${\mc O}(\ve)$. To ensure boundedness of solutions $\bPhi(x,t)$, we require the inhomogeneous portion of Eq.~($\ref{Leq}$) to be orthogonal to the nullspace of the adjoint operator $\mathcal{L^*}$. The nullspace of ${\mc L}^*$ is defined as the solution to the equation $\mathcal{L}^*\big[\bvphi(x)\big] = 0$, $\bvphi = (\varphi_1, \varphi_2, ..., \varphi_N)^T$, such that
\begin{align}
\label{eigfun}
\varphi_j(x) = f'(U_j(x)) \sum_{k=1}^N \int_{- \pi}^{\pi} w_{kj}(x-y)\varphi_k(y) \d y.
\end{align}
To derive explicit solutions to Eq.~(\ref{eigfun}), we must make further assumptions on either the firing rate function $f(u)$ or the weight functions $w_{jk}(x)$. For instance, if we assume symmetric interlaminar connectivity, such that $w_{jk}(x) \equiv w_{kj}(x)$ for all $j,k = 1,...,N$, then we can show that $\varphi_j(x) = f'(U_j(x))U'_j(x)$ (for all $j=1,...,N$) is a solution to Eq.~(\ref{eigfun}). This can be verified by applying integration by parts after plugging the expression into the integrand:
\begin{align*}
\varphi_j(x) &= f'(U_j) \sum_{k=1}^N w_{kj}(x)*\left[ f'(U_k(x))U'_k(x)\right] = f'(U_j) \sum_{k=1}^N w_{jk}(x)* \left[ f'(U_k(x))U'_k(x) \right] \\
& = f'(U_j) \frac{\d}{\d x} \sum_{k=1}^N w_{jk}(x)*f(U_k(x)) = f'(U_j(x)) U_j'(x), 
\end{align*}
where we have applied Eq.~(\ref{spatder}) in the last equality. Solutions can also be found for more general weight functions ($w_{jk}(x) \not\equiv w_{kj}(x)$), assuming $f(u) = H(u - \theta)$, the Heaviside nonlinearity, Eq.~(\ref{fheav}), as we demonstrate in Section \ref{sec:hfr}.

Assuming we can solve Eq.~(\ref{eigfun}), we enforce boundedness by taking the inner product of $\bvphi (x)$ with Eq.~(\ref{Leq}). For the time being, we assume the null space of ${\mc L}^*$ is one-dimensional, and address other cases in Section \ref{sec:ndb}. Thus, we take the null vector $\bvphi (x)$ and compute its inner product with the inhomogeneous portion of Eq.~(\ref{Leq}) to yield:
\begin{align*}
0 &= \sum_{j=1}^N \left\langle \varphi_j(x), \left[ {\mc L}_j \left[ \bPhi (x,t) \right] + \int_{- \pi}^{\pi} h_j(y+ \Delta) w_{jj}(x-y) f(U_j(y)) \d y  \right. \right. \\
& \hspace{30mm} \left. \left. + v(t) \sum_{k=1}^N \int_{- \pi}^{\pi} w_{vjk}(x-y) f(U_k(y)) \d y \right] \d t + \ve^{-1} \d \Delta U_j'(x) + \d Z_j(x,t)  \right\rangle.
\end{align*}
Rearranging terms leads to the following one-dimensional stochastic differential equation:
\begin{align}
\d \Delta (t) &= \left[ q(\Delta (t)) + \ve v(t) \right] \d t + \d {\mc Z}(t), \label{delsde}
\end{align}
where the terms on the right hand side include a weighted average of each layer's: (a) spatial heterogeneity $q(\Delta)$, (b) velocity $\ve v(t)$, and (c) noise ${\mc Z}(t)$. The impact of local spatial heterogeneity in each layer on the effective position $\Delta (t)$ is given by
\begin{align}
q( \Delta ) = \ve \Upsilon \left[ \sum_{j=1}^N \int_{- \pi}^{\pi} \varphi_j(x) \left( \int_{- \pi}^{\pi} h_j(y+ \Delta) w_{jj} (x-y) f(U_j(y)) \d y \right) \d x \right],  \label{qdelta}
\end{align}
where
\begin{align}
\Upsilon = -\left( \sum_{j=1}^N \mu_j \right)^{-1}, \hspace{5mm} \mu_j = \int_{- \pi}^{\pi} \varphi_j(x) U_j'(x) \d x, \hspace{5mm} j=1,...,N, \label{muj}
\end{align}
so in the absence of any velocity or noise, local attractors of the network are given by $\bar{\Delta}$ where $q( \bar{\Delta}) = 0$. Furthermore, the potential function, which determines statistical quantities such as mean first passage times, is given by $Q( \Delta) = - \int_{- \pi}^{\Delta} q(s) \d s$. Next, note that the effective velocity input to the multilayer bump solution is precisely $\ve v(t)$, which can be shown by applying our assumption on the weight functions $w_{vjk}(x)$, Eq.~(\ref{velwt}), to compute
\begin{align*}
\Upsilon \left[  v(t) \sum_{j=1}^N  \int_{- \pi}^{\pi} \varphi_j(x) \sum_{k=1}^N \int_{- \pi}^{\pi} w_{vjk}(x-y) f(U_k(y)) \d y \d x \right] = -\Upsilon \left( \sum_{j=1}^N \mu_j \right) v(t) = v(t), 
\end{align*}
where we have reduced the numerator by applying Eq.~(\ref{spatder}). Finally, the effective noise to the stochastic position variable $\Delta (t)$ is given by the spatially averaged and weighted process
\begin{align*}
\d {\mc Z}(t) = \ve \Upsilon \left[ \sum_{j=1}^N \int_{- \pi}^{\pi} \varphi_j (x) \d Z_j(x,t) \d x \right],
\end{align*}
which has zero mean $\langle {\mc Z}(t)  \rangle = 0$ and variance $\langle {\mc Z}(t)^2 \rangle = \bar{D} t$, where we can apply Eq.~(\ref{noiseeq}) for noise correlations to compute
\begin{align}
\bar{D} = \sum_{j=1}^N \sum_{j=1}^N D_{jk}, \hspace{8mm} D_{jk} = \ve^2 \Upsilon^2 \int_{- \pi}^{\pi} \int_{- \pi}^{\pi} \varphi_j(x) \varphi_k(y) C_{jk}(x-y) \d y \d x, \hspace{4mm} j,k=1,...,N, \label{diffcoeff}
\end{align}
demonstrating the contribution of the effective noise acting on $\Delta (t)$ will be determined by a weighted average of the noises from each layer $j=1,...,N$. To determine specific features of the dynamics of Eq.~(\ref{delsde}), we further define constituent functions of the model Eq.~(\ref{nfmodel}). To begin, we reduce the formulae considerably by focusing on the Heaviside nonlinearity, $f(u) = H(u - \theta)$, Eq.~(\ref{fheav}), allowing for analytic calculations of the above quantities.

\subsection{Results for a Heaviside firing rate}
\label{sec:hfr}
As demonstrated in Section \ref{blinstab}, assuming the firing rate function is a Heaviside nonlinearity, $f(u) = H(u- \theta)$, Eq.~(\ref{fheav}), can allow for direct calculation of eigensolutions to the linear stability problem for bumps ($\lambda \bpsi = {\mc L} \bpsi$). Identifying the nullspace of the adjoint linear operator is a related problem (${\mc L}^* \bvphi \equiv 0$), and assuming $f(u) = H(u-\theta)$ projects the infinite-dimensional problem to a $2N$-dimensional linear system. We then need only solve for a vector whose entries correspond to the coefficients of delta functions, as discussed in \cite{kilpatrick15}. To demonstrate, we first apply our formula for the derivative $f'(U_j(x)) = \gamma_j \left[ \delta(x-a_j) + \delta (x+a_j) \right]$, Eq.~(\ref{fprime}), and our formula for $\gamma_j$, Eq.~(\ref{gammak}). The delta functions contained in $f'(U_j(x))$ concentrate Eq.~(\ref{eigfun}) for $\varphi_j(x)$ at the set of $2N$ points, $x = \{ \pm a_1, \pm a_2, ..., \pm a_N \}$. This suggest the ansatz $\varphi_j(x) = \alpha_j \delta(x-a_j) + \beta_j \delta(x+a_j)$. Plugging these assumptions into Eq.~(\ref{eigfun}) reduces it to the form:
\begin{align*}
\varphi_j(x) & = \gamma_j \left[ \delta(x-a_j) + \delta(x+a_j) \right]\sum_{k=1}^N \int_{- \pi}^{\pi} w_{kj}(x-y) \left( \alpha_k \delta(y-a_j) + \beta_k \delta(y+a_j)\right) \d y  \\
&= \gamma_j  \sum_{k=1}^N \left[ \alpha_k w_{kj}(a_j - a_k) + \beta_k w_{kj}(a_j + a_k) \right] \delta (x-a_j) \\
& \hspace{8mm} + \gamma_j  \sum_{k=1}^N \left[ \alpha_k w_{kj}(a_j+ a_k) + \beta_k w_{kj}(a_j- a_k) \right] \delta (x+a_j), \hspace{4mm} j=1,...,N.
\end{align*}
Recalling that we have defined $\varphi_j(x) = \alpha_j \delta(x-a_j) + \beta_j \delta(x+a_j)$, we generate equations for the constants $\alpha_j$ and $\beta_j$ ($j=1,...,N$) by requiring self-consistency at $x = \{ \pm a_1, \pm a_2, ..., \pm a_N \}$:
\begin{align*}
\alpha_j = \gamma_j \sum_{k=1}^N \alpha_k w_{kj}(a_j - a_k) + \beta_k w_{kj}(a_j + a_k), \hspace{2mm} \beta_j = \gamma_j \sum_{k=1}^N \alpha_k w_{kj}(a_j - a_k) + \beta_k w_{kj}(a_j + a_k),
\end{align*}
for $j=1,...,N$, which can be expressed concisely as the $2N$-dimensional linear system:
\begin{align}
\z &= \W^* \z, \hspace{4mm} \z = \left( \begin{array}{c} \balpha \\  \bbeta \end{array} \right) =  (\alpha_1, ..., \alpha_N, \beta_1, ..., \beta_N)^T, \hspace{4mm} \W^* = \left[ \begin{array}{cc} {\mc A}_-^* & {\mc A}_+^* \\   {\mc A}_+^* & {\mc A}_-^* \end{array} \right],  \label{adjlins}
\end{align}
where $\W^*$ is the adjoint of the matrix defined in Eq.~(\ref{psieval}), the linear stability problem for stationary bumps. The system, Eq.~(\ref{adjlins}), can be written out in terms of its block matrix structure as
\begin{align}
\balpha = {\mc A}_-^* \balpha + {\mc A}_+^* \bbeta, \hspace{5mm} \bbeta = {\mc A}_-^* \bbeta + {\mc A}_+^* \balpha,  \label{block1}
\end{align}
which can be rearranged into the corresponding block matrix equations for $\balpha_{\pm} = \balpha \pm \bbeta$:
\begin{align}
\balpha_+ = \left( {\mc A}_-^*  + {\mc A}_+^* \right) \balpha_+, \hspace{5mm} \balpha_- = \left( {\mc A}_-^*  - {\mc A}_+^* \right) \balpha_-.  \label{block2}
\end{align}
For a nontrivial solution to Eq.~(\ref{block1}) to exist, there must be a nontrivial solution to either system in Eq.~(\ref{block2}). As demonstrated in Section \ref{blinstab}, there is always a nontrivial solution to $\mathbf{x} = \left( {\mc A}_-  - {\mc A}_+ \right) \mathbf{x}$, corresponding to the translation symmetric perturbation of the linear stability operator defined in Eq.~(\ref{psi3}). As this implies an eigenvalue of unity associated with $ \left( {\mc A}_-  - {\mc A}_+ \right)$, there must also be an eigenvalue of unity associated with $ \left( {\mc A}_-^*  - {\mc A}_+^* \right)$. In general, we do not expect nontrivial solutions to $\mathbf{x} = \left( {\mc A}_-  + {\mc A}_+ \right) \mathbf{x}$, and thus expect none for $\balpha_+ = \left( {\mc A}_-^*  + {\mc A}_+^* \right) \balpha_+$. This means, we expect $\balpha_+ \equiv \mathbf{0}$, so $\bbeta \equiv -\balpha$. Thus, we need only solve the $N$-dimensional system $\balpha = \left( {\mc A}_-^*  - {\mc A}_+^* \right) \balpha$. Applying these results to Eq.~(\ref{delsde}), we find a more tractable form for the integral terms defining each of the components:
\begin{align*}
q(\Delta) =  \ve \Upsilon \sum_{j=1}^N \alpha_j \int_{- a_j}^{a_j} h_j(y+\Delta) \left[ w_{jj}(a_j-y) - w_{jj}(a_j + y) \right] \d y,
\end{align*}
where now $\Upsilon = \left(2 \sum_{j=1}^{N} \alpha_j |U_j'(a_j)| \right)^{-1}$ for $j=1,...,N$, using the fact that $U'(-a_j) = -U_j'(a_j)>0$. Lastly, note the summed components of effective diffusion coefficient $\bar{D}$ are given by direct evaluations of the correlation functions
\begin{align*}
D_{jk} = 2 \ve^2 \Upsilon^2 \alpha_j \alpha_k \left[ C_{jk}(a_j - a_k) - C_{jk}(a_j + a_k) \right], \hspace{7mm} j,k=1,...,N,
\end{align*} 
reflecting the fact that noise primarily impacts the threshold crossing points of the bumps, initially at $x = \pm a_j$, $j=1,...,N$.

Mirroring our discussion in the linear stability Section \ref{blinstab}, we now discuss those cases with respect to the adjoint problem and note how they reflect the results derived there.

{\em Two-layer feedforward network.} Assuming $w_{12} \equiv 0$, layer 1 receives no feedback from layer 2. In this case, the coefficients $\alpha_{1}$ and $\alpha_2$ are given by
\begin{align}
\label{ffcoeff}
\alpha_1 = \frac{w_{11}^- \alpha_1 + w_{21}^- \alpha_2}{w_{11}^-}, \hspace{6mm} \alpha_2 = \frac{w_{22}^- \alpha_2}{w_{21}^- + w_{22}^-},
\end{align}
which has solutions $\alpha_2 = 0$ and $\alpha_1$ arbitrary, so the dynamics of the reduced system is entirely determined by those of layer 1. Layer 2 tracks the motion of the bump in layer 1, since $\alpha_2 = 0$: $\bar{D} = D_{11}$ and $q(\Delta) = - \ve \mu_1^{-1} \int_{- \pi}^{\pi} \varphi_1(x) \int_{- \pi}^{\pi} h_1(y+ \Delta) w_{11}(x-y) f(U_1(y)) \d y \d x$.

{\em Exploding star network.} For an arbitrary number of layers $N$, and $w_{jk} \equiv 0$ for $j=2,...,N$ and $k \neq j$, layer 1 receives no feedback from other layers and layers $2,...,N$ only receive input from layer 1.  In this case, the coefficients $\alpha_{j}$ are given
\begin{align}
\label{expcoeff}
\alpha_1 = \frac{\sum_{k=1}^N w_{k1}^- \alpha_k }{w_{11}^-}, \hspace{6mm} \alpha_j = \frac{w_{jj}^- \alpha_j}{w_{jj}^- + w_{j1}^-},
\end{align}
which has solutions $\alpha_j = 0$ for $j \neq 1$ and $\alpha_1$ arbitrary. All other layers track layer 1, thus the dynamics of the independent layer 1: $\bar{D} = D_{11}$ and $q(\Delta) = - \ve \mu_1^{-1} \int_{- \pi}^{\pi} \varphi_1(x) \int_{- \pi}^{\pi} h_1(y+ \Delta) w_{11}(x-y) f(U_1(y)) \d y \d x$. We shall treat the case of an imploding star in Section \ref{sec:ndb}. 

{\em Two-layer recurrent network.} In the case of a fully recurrent network, $\bar{w}_{jk} > 0$ for all $j \neq k$, we find the $N=2$ case yields the following set of equations for $\alpha_1$ and $\alpha_2$:
\begin{align*}
\alpha_1 = \frac{w_{11}^- \alpha_1}{w_{11}^- + w_{12}^-} + \frac{w_{21}^- \alpha_2}{w_{11}^- + w_{12}^-}, \hspace{6mm} \alpha_2 = \frac{w_{12}^- \alpha_1}{w_{22}^- + w_{21}^-} + \frac{w_{22}^- \alpha_2}{w_{22}^- + w_{21}^-},
\end{align*}
which can be reduced to the much simpler single equation, $w_{12}^- \alpha_1 = w_{21}^- \alpha_2$, so clearly $(\alpha_1, \alpha_2) = (w_{21}^-, w_{12}^-)$ is a solution as shown in \cite{kilpatrick15}. Thus, if $w_{21}(x)$ is the stronger connectivity function, then $\alpha_1$ will tend to be larger and layer 1 will have a larger influence on the overall dynamics. We see this clearly in the limiting feedforward case, in which $w_{12}(x) \equiv 0$.

{\em Directed loop of $N$ layers.} Lastly, we consider a directed loop of $N$ layers, wherein $w_{jk} \equiv 0$ unless $k=j$ or $k=j-1$ ($k=N$ for $j=1$). In this case, the equations for $\alpha_j$ are written
\begin{align*}
\alpha_j = \frac{w_{jj}^- \alpha_{j}}{w_{jj}^- + w_{j,j-1}^-} + \frac{w_{j+1,j}^- \alpha_{j+1}}{w_{jj}^- + w_{j,j-1}^-}, \hspace{4mm} j=1,...,N,
\end{align*}
where $j-1=N$ for $j=1$ and $j+1=1$ for $j=N$. Rearranging terms demonstrates that $w_{j,j-1}^- \alpha_j = w_{j+1,j}^- \alpha_{j+1}$, so $\alpha_j = 1/w_{j,j-1}^-$ ($j=1,...,N$) satisfies the system.

\section{Numerical simulations} 
\label{sec:scb}

\begin{figure}
\begin{center} \includegraphics[width=12cm]{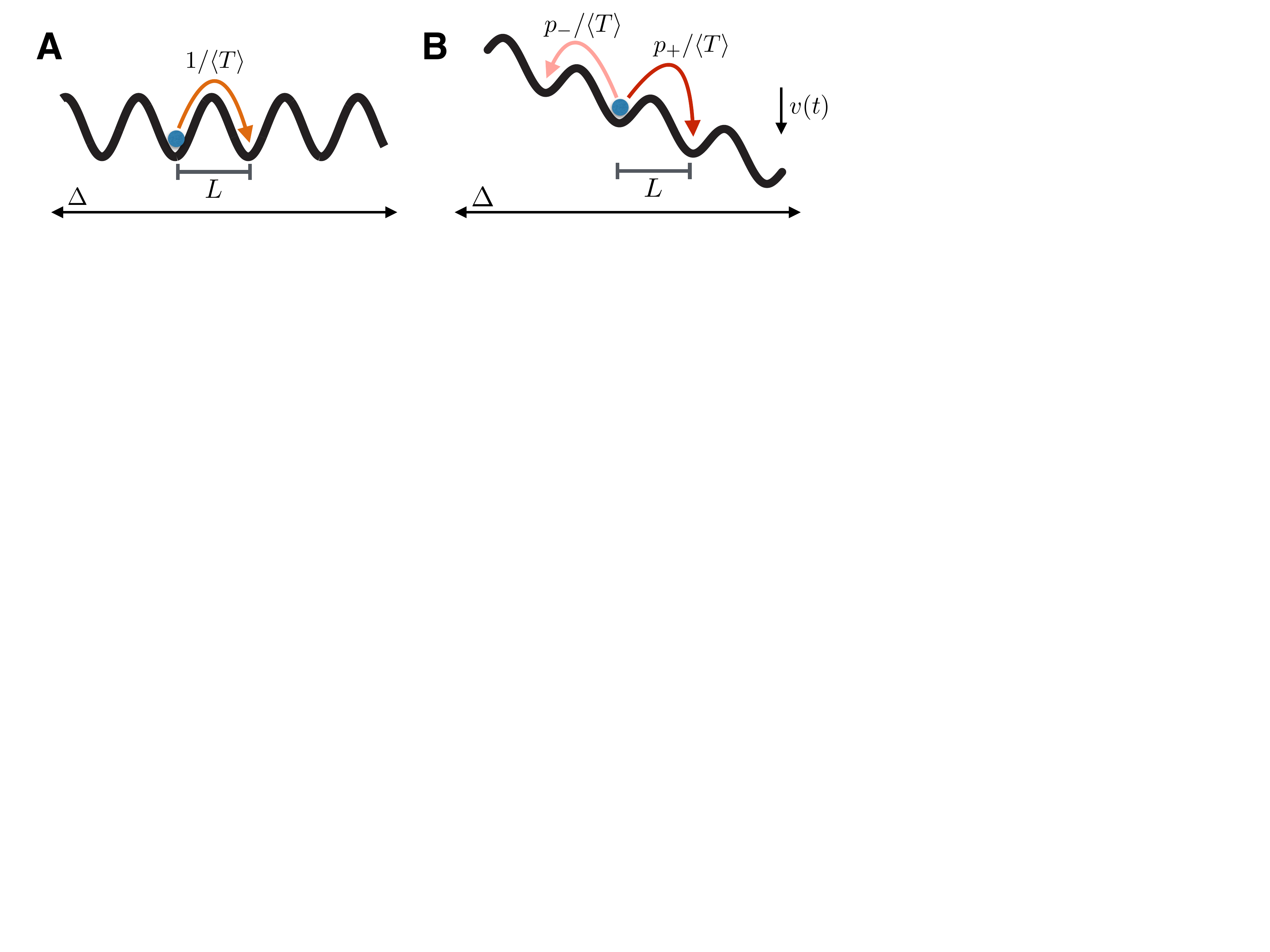} \end{center}
\caption{Effective diffusion and velocity calculations for the low-dimensional system, Eq.~(\ref{delsde}). For a gradient function $q(\Delta)$ with period $L$, we can determine the average ({\bf A}) diffusion of $\Delta(t)$ when $v(t) \equiv 0$ and ({\bf B}) velocity when $v(t) \neq 0$. ({\bf A}) The effective diffusion coefficient $D_{eff} = L^2/(2 \langle T \rangle)$ approximates the motion of the bump by tracking hops between neighboring potential wells that are distance $L$ apart, where average time between hops is $\langle T \rangle$~\cite{risken84,lindner01,kilpatrick13}. ({\bf B}) Velocity inputs tilt the potential $Q(\Delta)$, so, for example, the probability of a rightward transition is greater than a leftward one ($p_+ > p_-$). In this case, the bump has a nonzero effective velocity, approximated $v_{eff} = L(p_+-p_-)/ \langle T \rangle$~\cite{lindner01}.}
\label{fig6}
\end{figure}

In this section, we perform further analysis on Eq.~(\ref{delsde}) and compare with numerical simulations of Eq.~(\ref{nfmodel}). We are mainly interested in the interaction between noise and the spatial heterogeneity described by the nonlinear function $q(\Delta)$ in Eq.~(\ref{delsde}). In the absence of any velocity input, $v(t) \equiv 0$, we compute an effective diffusion coefficient $D_{\text{eff}}$, approximating the variance of $\Delta (t)$ given any periodic heterogeneity $q(\Delta)$ (Fig. \ref{fig6}A). In essence, we must compute the mean first passage time for trips between local attractors of Eq.~(\ref{delsde}). Velocity inputs subsequently tilt the potential determined by $Q(\Delta) = - \int_{- \pi}^{\Delta} \left[ q(s) + \ve v(t) \right] \d s$, so there is a bias in the direction of escapes from local attractors (Fig. \ref{fig6}B). Importantly, noise allows for propagation of bumps in instances where bumps would otherwise be stationary. We demonstrate the details of this analysis, and compare with simulations below.

\subsection{Specific multilayer architectures}
\label{specarch}
We now focus on specific examples of Eq.~(\ref{delsde}), where statistics of the resulting dynamics can be determined semi-analytically.

{\em Two-layer networks.} We begin by assuming $N=2$ with internal coupling is $w_{jj}(x) = \cos(x)$ with local heterogeneity $h_j(x) = \sigma_j \cos(n_jx)$ and interlaminar connectivity $w_{jk}(x) = \bar{w}_{jk} (1 + \cos x)/2$ ($j=1,2$, $k\neq j$). Note, this distinguishes this study from previous work in \cite{kilpatrick13b,kilpatrick15}, which assumed homogeneous connectivity.  We determined in Section \ref{sec:hfr} that $\alpha_1 = w_{21}^-$ and $\alpha_2=w_{12}^-$, allowing us to calculate the integrals in Eq.~(\ref{qdelta}) directly
\begin{align*}
q(\Delta) = -2 \ve \Upsilon \Big( w_{21}^- \mathcal{C}_1 \sin(a_1) \sin(n_1\Delta) + w_{12}^- \mathcal{C}_2 \sin(a_2) \sin(n_2\Delta) \Big)
\end{align*}
where $\Upsilon$ is given by Eq.~($\ref{muj}$) as 
\begin{align*}
\Upsilon = \left[ 2 \big( w_{21}^- \sin(a_1)( 2\sin (a_1) + \bar{w}_{12} \sin(a_2)) + w_{12}^- \sin(a_2)(2 \sin(a_2) + \bar{w}_{21} \sin(a_1) \big) \right]^{-1}
\end{align*}
and the impact of the heterogeneities scales like
\begin{align}
\mathcal{C}_j &= \sigma_j \dfrac{2 n_j \sin(a_j) \cos(n_j a_j) - 2 \cos(a_j) \sin(n_j a_j)}{n_j^2 - 1},  \hspace{7mm} j=1,2, \label{Cjform}
\end{align}
for $n_j \not = 1$. When $n_j = 1$, we may take the limit as $n_j \rightarrow 1$ of Eq.~($\ref{Cjform}$) so that $\mathcal{C}_j = \sigma_j \big(\sin(a_j)\cos(a_j) - a_j\big)/2$. Finally, we specify the spatial noise correlations as $C_{jj}(x) = \pi \cos (x)$ for $j=1,2$ and $C_{jk} (x) \equiv 0$ for $k \neq j$, so $D_{jk} \equiv 0$ for $k \neq j$ and the noise ${\mc Z}_t$ has diffusion coefficient
\begin{align*}
\bar{D} = D_{11}+D_{22} = 4 \ve^2 \Upsilon^2 \pi \left[ \left(w_{21}^-\right)^2 \sin^2(a_1) + \left(w_{12}^- \right)^2 \sin^2(a_2) \right].
\end{align*}
We now examine two specific cases of two-layer networks, simplifying these formulae further.

{\em Two-layer feedforward network.} In the case $\bar{w}_{12} = 0$, formulae for $\alpha_1$ and $\alpha_2$ are given in Eq.~(\ref{ffcoeff}), and without loss of generality we can set $\alpha_1 = 1$ and $\alpha_2 = 0$. Stochastic dynamics of the multilayer bump are thus approximated by the dynamics of the bump in layer 1, so the layer 2 bump tracks bump 1's position. The nonlinearity $q(\Delta) = - \ve \mathcal{C}_1 \sin (n_1\Delta)/(2 \sin(a_1))$ and the diffusion coefficient $\bar{D} = D_{11} = \pi \ve^2 /(4 \sin^2(a_1))$. Thus, the effective potential determining the bump's position $\Delta(t)$ is:
\begin{align*}
Q(\Delta) := -\int_{-\pi}^\Delta \left[q(s) + \ve v(t) \right] \d s = -\frac{\ve \mathcal{C}_1 \cos(n_1 \Delta(t))}{2n_1 \sin(a_1)} - \ve v(t)\Delta(t).
\end{align*}
We use this in determining the theoretical curves plotted in Figs. \ref{fig7} and \ref{fig8}, which we calculate in Section \ref{effdiff}. Essentially, we project the dynamics of Eq.~(\ref{delsde}) to a continuous-time Markov process whose transition rates are determined by the escape times from the local attractors, as illustrated in Fig. \ref{fig6}. 

{\em Two-layer symmetric network.} In the case $\bar{w}_{jk} = \bar{w}_c$, we have $a_j = a$ and $\alpha_j = 1$ ($j=1,2$, $k \neq j$), yielding $q(\Delta) = - \ve \left[ \mathcal{C}_1 \sin(n_1\Delta) + \mathcal{C}_2 \sin(n_2 \Delta) \right]/ \left[ (4 + 2 \bar{w}_c)\sin(a) \right]$ and $\bar{D} = \pi \ve^2/ \left[ 2 (2+ \bar{w}_c)^2 \sin^2 (a) \right]$. Note, the effective noise has diffusion coefficient that is substantially decreased as opposed to the single-layer or feedforward case~\cite{kilpatrick15}. Fluctuations are dampened by introducing loops in the connectivity of the multilayer network. Again, these functional forms are utilized in Figs. \ref{fig7} and \ref{fig8}.

{\em Exploding star network.} These results can also be generalized to $N$-layer networks that possess a valid one-dimensional projection described by Eq.~(\ref{delsde}). Another example is that of an exploding star, discussed in Section \ref{sec:hfr}. Assuming $w_{jk} \equiv 0$ for $j = 2,...,N$ and $k \neq j$, the coefficients, as computed in Eq.~(\ref{expcoeff}), are $\alpha_j = 0$ for $j > 1$ and $\alpha_1 = 1$. Thus, the dynamics of the stochastically-driven bump solution are determined by the dynamics of the independent layer 1, and other layers track these dynamics. Also, the constituent functions of the low-dimensional approximation will be exactly that of the two-layer feedforward example.

{\em Directed loop of $N$ layers.} Finally, we demonstrate the calculations for a directed loop of $N$ layers, wherein $w_{jk} \equiv 0$ unless $k=j$ or $k=j-1$ ($k=N$ for $j=1$). The coefficients $\alpha_j = 1/w_{j,j-1}^-$, as computed in Section \ref{sec:hfr}. Assuming $h_j(x) = \sigma_j \cos(n_jx)$, $w_{j,j-1}(x) = \bar{w}_{j,j-1} \big(1 + \cos(x)\big)/2$,  and $w_{jj}(x) = \cos(x)$, our low-dimensional approximation has form
\begin{align*}
q(\Delta) = -2 \ve \Upsilon \sum_{j=1}^N \frac{\mathcal{C}_{j} \sin(a_j)}{w_{j,j-1}^-} \sin(n_j \Delta), \hspace{7mm} \bar{D} = 4 \ve^2 \pi \Upsilon^2 \sum_{j=1}^N \left( \dfrac{ \sin(a_j)}{w_{j,j-1}^-} \right)^2,
\end{align*}
with $\Upsilon$ defined by Eq.~(\ref{muj}) and ${\mc C}_j = 2 \sigma_j \left[ n_j \sin(a_j) \cos(n_j a_j) - \cos(a_1) \sin(n_j a_j) \right]/(n_j^2 - 1)$ for $n_j \not = 1$ and $\mathcal{C}_j = \sigma_j \big(\sin(a_j)\cos(a_j) - a_j\big)/2$ for $n_j = 1$. Consider symmetry in the strength of synaptic connectivity, so that $\bar{w}_{j,j-1} = \bar{w}_c$ and $a_j = a$ for $j=1,...,N$, and 
\begin{align*}
\displaystyle q(\Delta) = - \ve \frac{\sum_{j=1}^N \mathcal{C}_{j} \sin (n_j \Delta)}{N(2+\bar{w}_c)\sin(a)} , \hspace{5mm} \bar{D} =  \frac{\ve^2 \pi}{N(2 + M_1)^2 \sin^2(a)}.
\end{align*}
We use these results in Fig. \ref{fig7}C.

With the constituent functions known, we analyze the stochastic differential equation to approximate the mean position $\langle \Delta (t) \rangle$ and variance $\langle \Delta^2(t) \rangle$ of the bump's position.

\subsection{Effective diffusion and velocity of the low-dimensional model} 
\label{effdiff}
Velocity integration must often be performed by spatial working memory networks involved in navigation or the head direction system~\cite{mcnaughton06,knierim12,geva15}. We consider the two main sources of error that could be incurred by a heterogeneous network subject to fluctuations. First, noise-driven diffusion of the remembered position will cause a degradation of spatial memory over time~\cite{compte00,kilpatrick13c}. Second, heterogeneities will lead to erroneous integration of the velocity inputs, since the network will not integrate them perfectly~\cite{seung96,brody03}. Thus, errors made in encoding the true position will arise from the noise term $\d \mathcal{Z}_t$ in Eq.~($\ref{delsde}$) as well as the heterogeneity $q(\Delta)$, so $ \d \mathcal{Z}_t \equiv 0$ and $q(\Delta) \equiv 0$ would yield perfect integration. We can asymptotically quantify these contributions to error by approximating (a) the effective diffusion: $\langle \Delta^2(t) \rangle - \langle \Delta (t) \rangle^2  \approx D_{eff}t$, and (b) the effective velocity: $\langle \Delta (t) \rangle \approx v_{eff} t$.

\begin{figure}
\begin{center} \includegraphics[width=16cm]{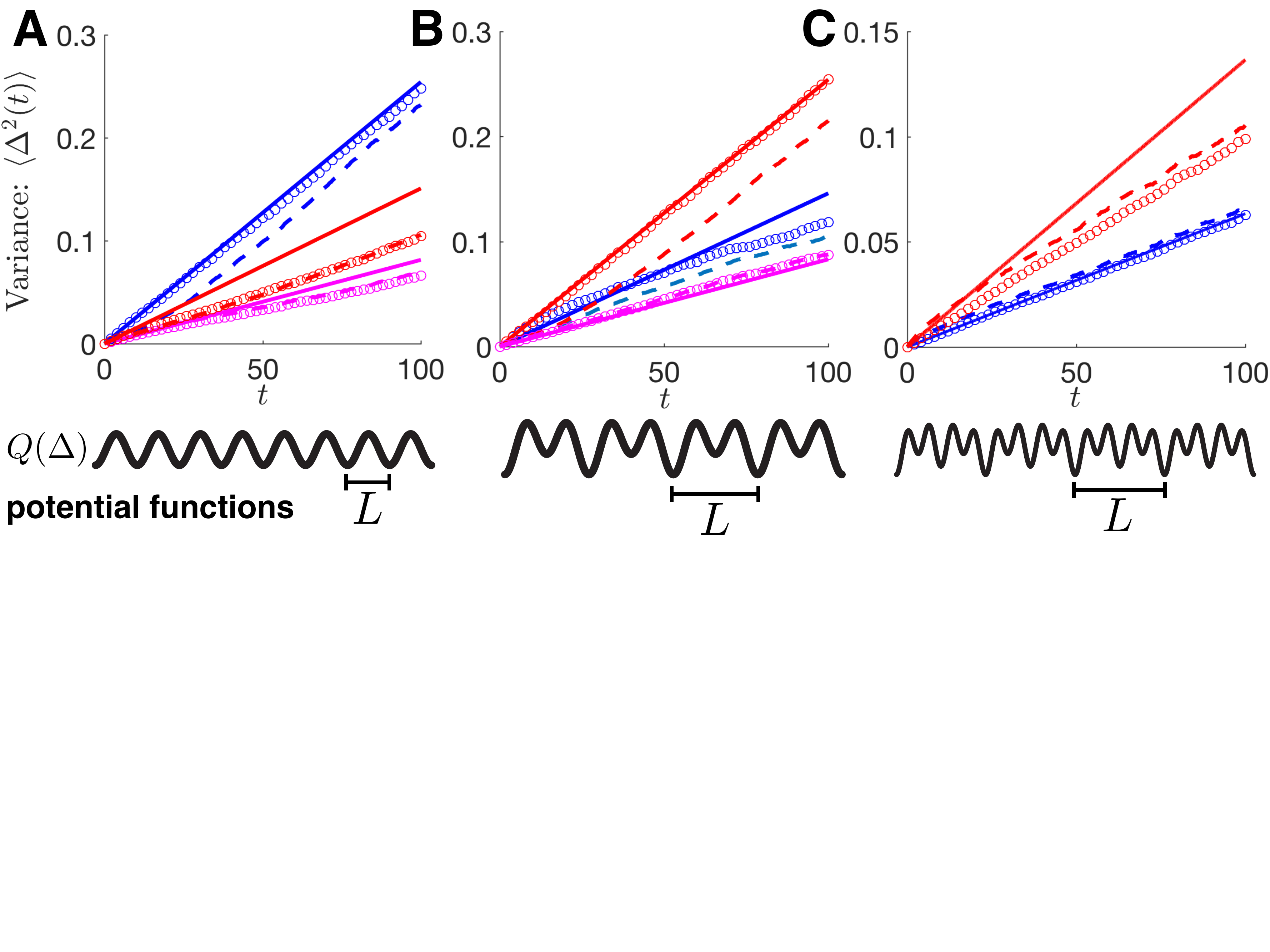} \end{center}
\caption{Variance $\langle \Delta^2 (t) \rangle$ of the bump solutions in the absence of a velocity input ($v(t) \equiv 0$) for fixed heterogeneities $h_j(x) = \sigma_j \cos (n_j x)$ ($j=1,2$) and varied interlaminar connectivity strengths $\bar{w}_{12}$ and $\bar{w}_{21}$. Qualitative descriptions of the associated potential functions $Q(\Delta)$ of each network are plotted below each panel. ({\bf A}) Plots in the case of symmetric heterogeneity ($n_1=n_2= 8$; $\sigma_1 = \sigma_2 = 0.25$) with feedforward connectivity (blue curves: $\bar{w}_{12} = 0.3, \bar{w}_{21} = 0$), asymmetric connectivity (red curves: $\bar{w}_{12} = 0.3, \bar{w}_{21} = 0.1$), and symmetric connectivity (magenta curves: $\bar{w}_{12} = \bar{w}_{21} = 0.3$). We find that statistics calculated from numerical simulations of the low-dimensional system (circles), Eq.~(\ref{delsde}) are well matched to statistics of simulations of the full model (dashed line), Eq.~(\ref{nfmodel}). Variances approximated by our effective diffusion calculation $\langle \Delta^2(t) \rangle = D_{eff} t$, Eq.~(\ref{deff}), are given by solid lines. Note, as interlaminar connectivity increases in strength, the variance scales more slowly with time. ({\bf B}) Plots in the case of asymmetric heterogeneity ($n_1=4$, $n_2=8$; $\sigma_1 = 0.05$, $\sigma_2=0.25$) where the the low-frequency ($n_1=4$), more stable layer determines dynamics (blue curves: $\bar{w}_{12} = 0.3, \bar{w}_{21} = 0$);  high-frequency ($n_2=8$), less stable layer determines dynamics (red curves: $\bar{w}_{12} = 0, \bar{w}_{21} = 0.3$); and symmetric coupling (magenta curves: $\bar{w}_{12} = \bar{w}_{21} = 0.3$). ({\bf C}) Plots for a system with $N=3$ layers where the heterogeneity: $h_1(x) = 0.01 \cos (4x)$, $h_2(x) = 0.025 \cos (8x)$, $h_3(x) = 0.25 \cos (16x)$. Connectivity is taken to be feedforward (red curves: $\bar{w}_{21} = \bar{w}_{32} = 0.3$, $\bar{w}_{jk} =0$ for all other $k \neq j$) and a symmetric loop (blue curves: $\bar{w}_{13} = \bar{w}_{21} = \bar{w}_{32} = 0.3$, $\bar{w}_{jk} =0$ for all other $k \neq j$). In all panels, $\ve = 0.1$. Numerical simulations of the full model, Eq.~(\ref{nfmodel}), were performed using Euler-Maruyama with timestep $dt=0.01$ with direct integration of convolution using $dx=0.01$ and $10^6$ realizations to compute ensemble statistics.}
\label{fig7}
\end{figure}

{\em Effective diffusion.} To compare with our results from full numerical simulations, we begin by deriving the effective diffusion coefficient of a bump evolving in a spatially heterogeneous network. This leverages previous results on transport in periodic potentials~\cite{risken84,lindner01}.  In the absence of velocity inputs, $v(t) \equiv 0$, we can approximate the stochastic motion of a bump by tracking the nearest positional attractor to its vicinity~\cite{kilpatrick13,kilpatrick13c}. Given a gradient function $q(\Delta)$ in Eq.~(\ref{delsde}), attractors $\bar{\Delta}$ obey $q(\bar{\Delta}) = 0$. For instance gradient functions of the form $q(\Delta) = -|K| \sin (n \Delta)$ have stable (unstable) attractors at $\bar{\Delta}_s = \frac{2 j \pi}{n} $ ($\bar{\Delta}_u = \frac{(2 j+1) \pi}{n} $). In our network, the distance $x_s$ between two stable attractors may not be equal to the period $L$ of the gradient function ($q(\Delta) = q(\Delta+L)$). In this case, we can either: (a) construct the corresponding continuous-time Markov chain model and compute the stochastic motion as such or (b) use a first passage time calculation to determine the mean time $\langle T \rangle$ until the bump evolves one period $L$ and use this in the standard effective diffusion calculation. We opt for the latter, so to begin, we note the general form of the effective diffusion coefficient (Details of the derivation can be found in \cite{lindner01,kilpatrick13,kilpatrick13c}):
\begin{align}
\label{deff}
\displaystyle D_{eff} = \frac{\bar{D} \cdot L^2}{\langle T \rangle} = \frac{\bar{D} \cdot L^2 }{\displaystyle \int_0^{L} \int_0^L \e^{Q(x) - Q(y)} \d y \d x},
\end{align}
where $Q(\Delta) = - \int_{- \pi}^{\Delta} q(s) \d s$ is the potential given by integrating the gradient function as such. In the case of gradient functions $q(\Delta) = \sum_{j=1}^{N} K_j \sin (n_j \Delta)$, the period of the potential will be $L = (2 \pi)/n_{min}$ where $n_{min} = \min \{n_1,...,n_N\}$. Integrals in Eq.~(\ref{deff}) arising from simple trigonometric potential functions like $Q(\Delta) = \kappa \cos (n \Delta)$ can be expressed in terms of modified Bessel functions~\cite{kilpatrick13,kilpatrick13c}. However, the mixed mode potentials of interest do not yield integrals that can be evaluated analytically. Thus, for our comparisons with numerical simulations in Figs. \ref{fig7} and \ref{fig9}, we simply evaluate these integrals using numerical quadrature.

We find that the asymptotic approximation $\langle \Delta^2 (t) \rangle \approx D_{eff} t$ captures the trends in numerical simulations reasonably well. In Fig. \ref{fig7}A, we analyze the diffusion of bumps in a multilayer network with the same spatial heterogeneity function in each layer ($h_1(x) \equiv h_2(x)$). As in previous work~\cite{kilpatrick13b}, increasing the strength of interlaminar connectivity decreases the rate at which the variance scales in time. Furthermore, the purely feedforward network has far higher variance than a network with weakly recurrent coupling, since the network bump position is controlled by a single layer. As a result, the noise cancelation that arises from recurrent coupling is not apparent. In Fig. \ref{fig7}B, we study the effects of having two layers with different spatial heterogeneity ($h_1(x) = \sigma_1 \cos (4x)$, $h_2(x) = \sigma_2 \cos (8x)$). Note the multimodal shape of the effective potential $Q(\Delta)$. As a result, different feedforward architectures ($1\mapsto2$ vs. $2\mapsto1$) can lead to substantially different variances, depending on whether the more stable layer 1 or less stable layer 2 determines the dynamics. Lower frequency spatial heterogeneities tend to stabilize bumps more to stochastic perturbations, generally leading to a lower effective diffusion~\cite{kilpatrick13,kilpatrick13c}. Here, we show that this feature influences which interlaminar coupling architectures are best for reducing variance in spatial working memory. Lastly, we study a three-layer network in Fig. \ref{fig7}C. A fully recurrent architecture reduces the diffusion of the bump more than a feedforward architecture, even when the feedforward architecture is dominated by the layers that are more robust to noise perturbations ($h_1(x) = \sigma_1 \cos (4x)$, $h_2(x) = \sigma_2 \cos (8x)$). When interlaminar coupling from the less stable layer is incorporated ($h_3(x) = \sigma_3 \cos (16x)$), variance drops. Having validated our theory of effective diffusion for networks without velocity inputs, we now study the interaction of velocity inputs, noise, and spatial heterogeneity in multilayer networks.

{\em Effective velocity.} We now explore the impact of noise and spatial heterogeneity on the integration of velocity. While an analogous formula for the effective diffusion could also be derived, the results are quite similar to the case of no velocity inputs discussed above. Thus, we primarily consider how noise and heterogeneity contribute to the integration of velocity, as this will also be the main source of error when the network is integrating velocity. 

\begin{figure}
\begin{center} \includegraphics[width=12cm]{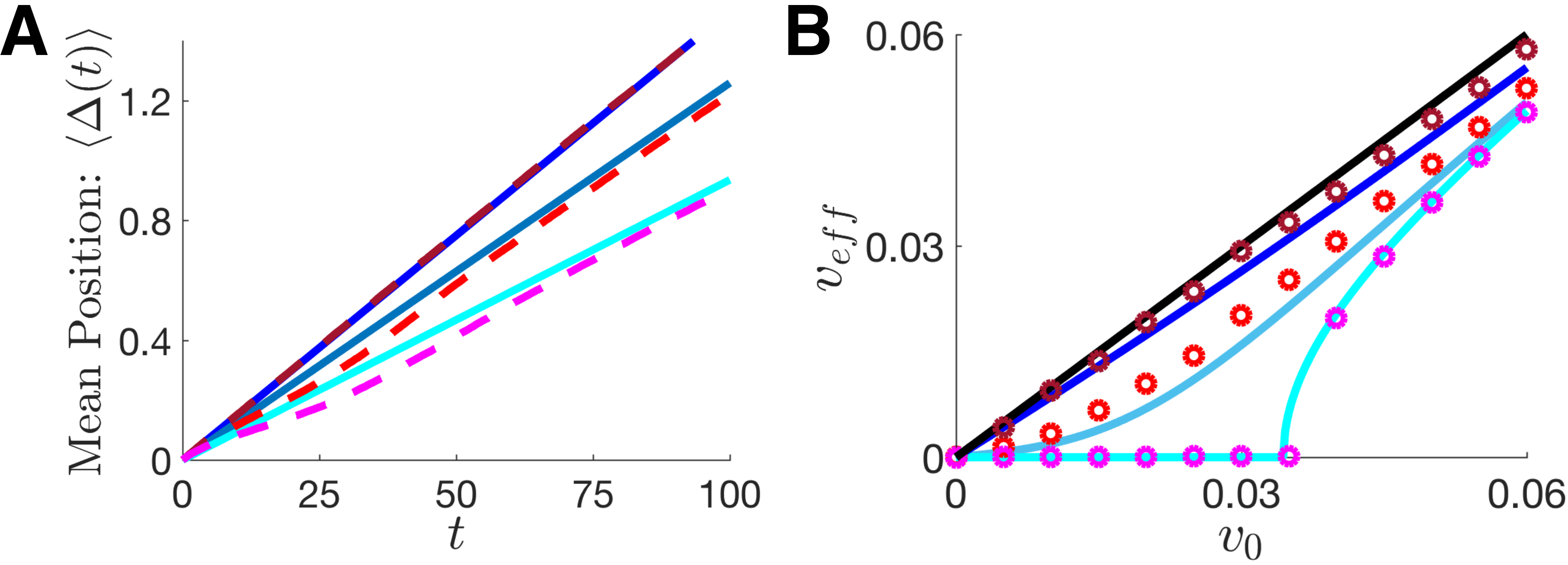} \end{center}
\caption{Bumps driven by velocity inputs ($\ve v(t) \equiv v_0 > 0$) impacted by spatial heterogeneities ($h_1(x) = \sigma_1 \cos (n_1x)$, $h_2(x) = \sigma_2 \cos (n_2 x)$) and noise. ({\bf A}) Plots of the mean bump position $\langle \Delta (t) \rangle$ for homogeneous networks (top lines: $\sigma_1=\sigma_2 = 0$), identical heterogeneity (middle lines: $n_1=n_2=16$; $\sigma_1=\sigma_2 = 1$), and differing heterogeneity (bottom lines: $n_1 = 8, n_2 = 16$; $\sigma_1 = \sigma_2 = 1$) in the layers. Note the top lines represent perfect integration of the $v_0 = 0.015$ velocity input in the ensemble average $\langle \Delta (t) \rangle$, whereas incorporating heterogeneity slows the propagation of bumps, so the velocity is integrated imperfectly. The theoretical lines (solid) computed from Eq.~(\ref{veleff}) match the results of numerical simulations (dashed lines) quite well.  ({\bf B}) Plot of the effective velocity $v_{eff}$ of the ensemble versus the input velocity $v_0$ as the noise strength is varied (bottom to top: $\ve = 0, 0.1, 0.2$. We fix the heterogeneity so that $n_1 = n_2 = 8$ in all curves, and $\ve \sigma_j = 0.4$, so that even in the limit of no noise ($\ve \to 0$), there is spatial heterogeneity. In the absence of noise, heterogeneity causes the bump to become pinned for sufficiently small velocity input $v_0$. Introducing noise causes the average effective velocity $v_{eff} = \langle \Delta (t) \rangle/t$ to approach the input velocity $v_0$. Blue solid lines are from theory Eq.~(\ref{veleff}), and circles are from numerical simulations. Black line is $v_{eff} = v_0$. For both panels, the coupling strength is symmetric: $\bar{w}_{12} = \bar{w}_{21} = 0.3$. Numerical simulations of the full model (dashed lines) are performed as described in Fig. \ref{fig7}.}
\label{fig8}
\end{figure}

Consider a velocity function $v(t)$ that is piecewise constant in time ($v(t) \equiv v_j$ on $t_j<t<t_{j+1}$), corresponding to the saltatory motion common to foraging animals~\cite{obrien90}. In this case, we can approximate the effective velocity $v_{eff}$ of bumps in the spatial working memory network, Eq.~(\ref{nfmodel}), by again computing the mean time of a transit of the variable $\Delta (t)$ across one period $L$ of the potential $Q(\Delta)$. We slightly abuse the notion of a period, since the velocity input $v_j$ will skew the potential as $Q(\Delta) = - \int_{- \pi}^{\Delta} q(s) \d s - \ve v_j \Delta$, so really $L$ represents the period of the $q(\Delta)$ portion of the potential. Our approximation proceeds by tracking the expected number of hops the bump makes. Hops occur when the bump leaves the vicinity of its local attractor and arrives in the vicinity of a neighboring attractor, presumably a distance $L$ away. Note, for multimodal potentials, we must account for the multiple attractors in a single period $L$, but we forgo those details here. Hops can be rightward $\chi_+(t)$ or leftward $\chi_-(t)$, so we track the difference $\chi (t) : = \chi_+(t) - \chi_-(t)$ to determine the rightward displacement.  Shifting coordinates to assume the bump begins at $\Delta (0) = 0$, we can approximate the position of the bump $\Delta (t) = L \cdot \chi (t)$. Since the counting process $\chi (t)$ is Markovian, we need only know the hop rates $p_{\pm}/\langle T \rangle$ to compute $\langle \chi (t) \rangle = \left[ p_+ - p_- \right] t / \langle T \rangle$, where $p_{+}$ ($p_-$) is the probability of a rightward (leftward) hop. The escape probabilities $p_{\pm}$, mean escape time $\langle T \rangle$, and effective velocity of the bump $v_{eff} = L \cdot \chi (t) \rangle / t$ can be calculated directly from Eq.~(\ref{delsde}) with the potential $Q(\Delta)$:
\begin{align}
v_{eff} = \frac{L(p_+-p_-)}{\langle T\rangle}, \hspace{2mm} p_+=1-p_- = \frac{1}{1 + \e^{-v_0L/\bar{D}}}, \hspace{2mm} \langle T \rangle = \frac{p_+}{\bar{D}} \int_0^L \int_{x-L}^x \e^{\frac{Q(x) - Q(y)}{\bar{D}}} \d y \d x.  \label{veleff}
\end{align}
We compare our formula for the effective velocity, Eq.~(\ref{veleff}), to results from numerical simulations in Fig. \ref{fig8}A. As the amplitude of heterogeneity increases, the effective speed of traveling bumps decreases, given identical velocity input. This is in line with previous studies on the impact of heterogeneities on wave propagation~\cite{bressloff01,poll16}. However, we also show that as the amplitude of noise is increased, the effective velocity $v_{eff}$ gets closer to $v_0$ (Fig. \ref{fig8}B). This is due to the fact that noise-induced transitions between local attractors become more frequent, and the motion of the bump reflects the asymmetry in the potential $Q(\Delta)$. In the case of large amplitude noise, $\bar{D} \gg 1$, we can approximate the transition probabilities and mean first exit time in Eq.~(\ref{veleff}) using linearization in the small parameter $1/ \bar{D}$: $\displaystyle p_+ \approx \frac{1}{2} + \frac{v_0L}{4D}$ and $\displaystyle \langle T \rangle \approx \frac{L^2}{2 \bar{D}}$, yielding $v_{eff} \approx v_0$. Thus, while the effective diffusion will also tend to increase with $\bar{D}$, the effective velocity will grow to more closely match the true input velocity, similar to results discussed in the optimal transport framework of \cite{lindner01}.

In the absence of noise ($\bar{D} \to 0$), we can no longer assume the bump stochastically transitions between local attractors. In fact, for persistent propagation in the network to occur, the gradient function $q(\Delta)$ must have no zeroes, so that $\dot{\Delta}(t) = q(\Delta) >0$ for all $\Delta$, assuming $v_0 > 0$. In this case, we can compute the time $\displaystyle T_L = \int_0^L \frac{\d \Delta}{q(\Delta)}$ it takes to traverse a single period $L$, and compute the effective velocity: $v_{eff} = L/T_L$ (For more details, see\cite{poll16}). For example, when $\dot{\Delta}(t) = - K \sin (n \Delta) + v_0$, the time it takes to traverse the length $L = 2 \pi/n$ is $T_L = 2 \pi / \left[m \sqrt{v_0^2 - K^2} \right]$ so $v_{eff} = 2 \pi / (mT) = \sqrt{v_0^2 - K^2}$. Clearly, if $v_0 \leq K$, this theory predicts the bump becomes pinned to a local attractor of the network, due to the spatial heterogeneity. The lower curve in Fig. \ref{fig8}B compares this theory with simulations of the noise-free version of the model Eq.~(\ref{nfmodel}), and indeed we find that heterogeneities then pin bumps so that $v_{eff} = 0$, in the absence of noise. On the other hand, the bump propagates in the presence of noise, so the velocity signal is detectable whereas it would not be in a noise-free paradigm, providing an example of stochastic resonance~\cite{longtin93}.

We conclude that, not only does our low-dimensional approximation describe bump dynamics in a multilayer network, it provides further insight into how heterogeneity, noise, and interlaminar coupling impact the encoding of input signals. Noise degrades positional information, but strong spatial heterogeneity and interlaminar coupling can stabilize bump positions over long delay periods. While heterogeneity disrupts integration of velocity inputs, sufficiently strong noise can restore mean bump propagation speeds to be close to the input velocity. Thus, there is a tradeoff between the stabilizing effects of heterogeneity and the resulting disruption of velocity integration, which we shall explore more in future work.
 
\section{Networks with multiple independent modules}
\label{sec:ndb}
\begin{figure}
\begin{center} \includegraphics[width=14cm]{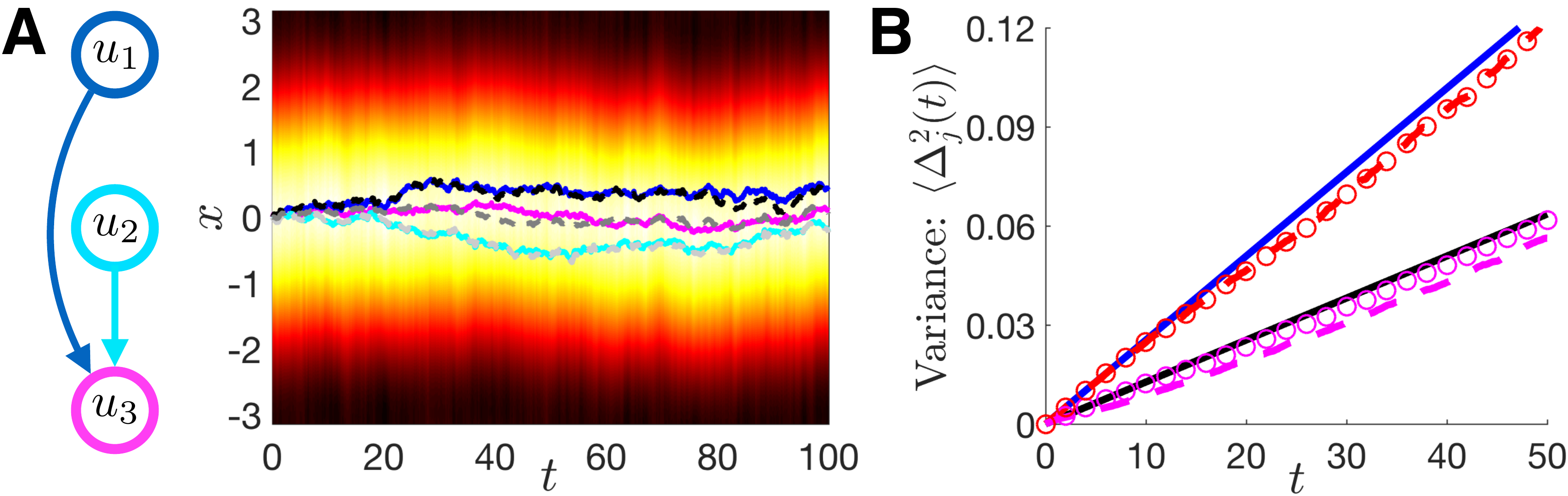} \end{center}
\caption{Evolution of bump positions in a $N=3$-layer network with only feedforward connectivity from $1\mapsto3$ and $2\mapsto3$, so layers 1 and 2 evolve independently. ({\bf A}) Numerical simulation of bump evolution in layer 3 overlaid with bump position from full simulation (magenta line), as well as positions of the bumps in layer 1 (dark blue) and layer 2 (cyan). Dashed lines are approximations from low-dimensional system, Eq.~(\ref{delsdej}). ({\bf B}) Variance $\langle \Delta_j^2(t) \rangle$ as a function of time as computed from full numerical simulations (dashed lines), the low-dimensional approximation (circles), and the effective diffusion calculation (solid lines), Eq.~(\ref{deff}). The top curves are for layer 1 ($\langle \Delta_1^2(t) \rangle$) and the bottom curves are for layer 3 ($\langle \Delta_3^2(t) \rangle$), the output layer. While this is a fully feedforward network, the output layer averages the position estimates in layers 1 and 2, reducing the effective diffusion of the layer 3 bump. Here $h_1(x) = h_2(x) = 0.25 \cos (8x)$ and $h_3(x) \equiv 0$ with interlaminar connectivity $\bar{w}_{31} = \bar{w}_{32} = 0.3$, and $\ve = 0.1$. Numerical simulations are performed as described in Fig. \ref{fig7}.}
\label{fig9}
\end{figure}

Our reduction to the low-dimensional system carried out in Section \ref{dereff} relied on the assumption that the multilayer bump solution possessed one marginally stable mode of perturbation. Thus, noise and velocity perturbations were always effectively integrated by the bumps in each layer by the same amount, so the multilayer bump moved coherently. However, if the interlaminar weight functions $w_{jk}$ of the network Eq.~(\ref{nfmodel}) are defined such that multiple layers receive no feedback from other layers, those independent layers only integrate perturbations of their own activity. Consider the three-layer imploding star network presented in Fig. \ref{fig5}: Shifting the bump in layer 1 does not impact the dynamics of the layer 2 bump. Thus, only noise perturbations local to those layers impact their activity (Fig. \ref{fig9}). This suggests we need to modify our derivation of a low-dimensional system to account for this independence.

This idea can be applied to a class of cases wherein $w_{jk} \equiv 0$ for $j=1,...,M$, where $M \leq N$, and $k \neq j$. With this assumption, we must now assume each bump in each layer $j=1,...,M$ has an independent phase $\Delta_j$. Subsequently, the remaining phases $\Delta_k$ for $k=M+1,...,N$ depend on the first $M$ phases. Since layers $1,...,M$ dominate the dynamics, we ignore the impacts of heterogeneity in layers $M+1,...,N$, so $h_j(x) \equiv 0$ for $j=M+1,...,N$. To begin, we consider the ansatz $u_j(x,t) = U_j(x-\Delta_j(t)) + \epsilon \Phi_j(x-\Delta_j(t),t)$ for all $j=1,...,N$. Plugging this into Eq.~($\ref{nfmodel}$) and truncating to $\mathcal{O}(\epsilon)$, we have:
\begin{subequations} \label{starLeq}
\begin{align}
\d \Phi_j =& \left[  \mathcal{L}_j\big[ \bPhi\big]  + w_{jj}(x)*\left[ f(U_j(x)) h_j(x+\Delta_j) \right]+ v(t) \sum_{k=1}^N w_{vjk}*f(U_k) \right] \d t \nonumber \\
& \hspace{7mm} + \ve^{-1} \d \Delta_j U'_j+  \d Z_j, \hspace{5mm} j=1,...,M,  \\
\d \Phi_j =& \left[  \mathcal{L}_j\left[ \bPhi \right] + v(t) \sum_{k=1}^N w_{vjk}*f(U_k) + \sum_{k \neq j} w_{jk}* \left[f'(U_k)U_k' \right] (\Delta_j - \Delta_k)  \right] \d t \nonumber \\
& \hspace{2mm} + \ve^{-1} \d \Delta_j U'_j+  \d Z_j,  \hspace{5mm} j = M+1,...,N, 
\end{align}
\end{subequations}
where we have linearized the terms $f(U_j(x+\Delta_j-\Delta_k)) = f(U_j(x)) + f'(U_j(x)) U_j'(x) (\Delta_j - \Delta_k)$ and recall $F(x)*G(x) = \int_{- \pi}^{\pi} F(x-y) G(y) \d y $. While the linearization in $(\Delta_j - \Delta_k)$ assumes the quantity remains small, our approximation performs reasonably well, even when bumps are substantially separated in numerical simulations (Fig. \ref{fig9}).  Note, $\mathcal{L}_j$ is the $j^{th}$ element of the linear functional $\mathcal{L}: \mathbf{p} \mapsto \mathbf{q}$ for $\mathbf{p}=(p_1,p_2,...,p_N)^T$ and $\mathbf{q}=(q_1,q_2,...,q_N)^T$, defined as
\begin{align*}
\mathcal{L}_j\big[ \mathbf{p}(x) \big] &= -p_j(x) + w_{jj}(x)*\left[f'(U_j(x))p_k(x) \right], \hspace{15mm} j=1,...,M, \\
\mathcal{L}_j\big[ \mathbf{p}(x) \big] &= -p_j(x) + \sum_{k=1}^N w_{jk}(x)*\left[ f'(U_k(x))p_k(x)\right], \hspace{8mm} j=M+1,...,N,
\end{align*}
with adjoint operator $\mathcal{L}^*: \mathbf{q} \mapsto \mathbf{p}$, defined $\langle {\mc L} \mathbf{p}, \mathbf{q} \rangle = \langle \mathbf{p}, {\mc L}^*\mathbf{q} \rangle $ under the standard $L^2$ inner product, and thus given element-wise by given for $j=1,...,N,$
\begin{equation*}
\mathcal{L}^*_j\big[ \mathbf{q}(x)\big] = -q_j(x) + f'(U_j(x)) \left[ w_{jj}(x)*q_j(x) + \sum_{k \neq j; \ k=M+1}^N w_{kj}(x)*q_k(x) \right].
\end{equation*}
To ensure boundedness of solutions $\bPhi(x,t)$, we require the inhomogeneous portion of Eq.~(\ref{starLeq}) to be orthogonal to the nullspace of the adjoint operator $\mathcal{L^*}$. Vectors $\bvphi = (\varphi_1, \varphi_2, ..., \varphi_N)^T$ that reside in the nullspace of ${\mc L}^*$ are solutions to the equation $\mathcal{L}^*\big[\bvphi(x)\big] = 0$, such that
\begin{align}
\label{eigfunstar}
\varphi_j(x) = f'(U_j(x)) \left[ w_{jj}(x)*\varphi_j(x) + \sum_{k \neq j; \ k=M+1}^N w_{kj}(x)*\varphi_k(x) \right], \hspace{5mm} j=1,...,N.
\end{align}
Solutions of Eq.~(\ref{eigfunstar}) can be identified by recalling the formula for the spatial derivative of $U_j(x)$, given by Eq.~(\ref{spatder}), and noting that for $j=1,...,N$, we have
\begin{align*}
U_j'(x) = \int_{- \pi}^{\pi} \frac{\d}{\d x} w_{jj}(x-y) f(U_j(y)) \d y = \int_{- \pi}^{\pi} w_{jj}(x-y) f'(U_j(y))U_j'(y) \d y.
\end{align*}
Therefore, if we set $\varphi_j(x) = f'(U_j(x))U_j'(x)$ for a {\em single index} $j=1,...,M$ and $\varphi_l(x) \equiv 0$ otherwise, then for that index $j$, Eq.~(\ref{eigfunstar}) becomes
\begin{align*}
f'(U_j(x))U_j'(x) = f'(U_j(x)) \left[ w_{jj}(x)*\left[ f'(U_j(x))U_j'(x) \right] + \sum_{k=M+1}^N w_{kj}(x)*\left( 0 \right) \right] = f'(U_j(x)) U_j'(x),
\end{align*}
and for $l \neq j$, we have
\begin{align*}
0 = f'(U_l(x)) \left[ w_{ll}(x)*\left[ 0\right] + \sum_{k \neq l; \ k=M+1}^N w_{kl}(x)*\left( 0 \right) \right] = 0.
\end{align*}
Thus, taking the inner product of Eq.~(\ref{starLeq}) with each function in this $M$-dimensional set of nullspace vectors, we have a closed system of independent evolution equations for the set of phases $(\Delta_1, ..., \Delta_M)$:
\begin{align}
\d \Delta_j = \left[ q_j(\Delta_j) + \ve v(t) \right] \d t + \d {\mc Z}_t^j, \hspace{5mm} j=1,...,M,  \label{delsdej}
\end{align}
where now $\langle \left( {\mc Z}_t^j \right)^2 \rangle = D_{jj}t$ with
\begin{align*}
q_j(\Delta_j) &= \ve \dfrac{ \int_{-\pi}^\pi f'(U_j(x))U_j'(x) \int_{-\pi}^\pi h_j(y + \Delta_j)w_{jj}(x-y)f(U_j(y)) \d y \d x}{\int_{-\pi}^\pi f'(U_j(x)) U'_j(x)^2 \d x},  \\
D_{jj} &= \ve^2 \frac{\int_{- \pi}^{\pi} \int_{- \pi}^{\pi} f'(U_j(x))U_j'(x) f'(U_j(y))U_j'(y) C_{jj}(x-y) \d y \d x}{\left[ \int_{- \pi}^{\pi} f'(U_j(x))U_j'(x)^2 \d x \right]^2}.
\end{align*}
Lastly, to express the phases $(\Delta_{M+1},...,\Delta_N)$ in terms of $(\Delta_1,...,\Delta_M)$, we apply the eigenvalue equation, Eq.~(\ref{psi3}), we derived in Section \ref{blinstab}. The possible equilibrium positions $(\Delta_1,...,\Delta_N)$ of bumps can be approximated by assuming a zero eigenvalue $\lambda = 0$ in Eq.~(\ref{psi3}) and taking inner products with $f'(U_j(x))U_j'(x)$ for $j=1,...,N$:
\begin{align}
\Delta_j \langle f'(U_j)U_j', U_j' \rangle &= \left\langle f'(U_j)U_j' ,  \sum_{k=1}^N w_{jk}*\left[f'(U_k) U_k' \right] \Delta_k \right\rangle, \nonumber \\
0 &= \left\langle f'(U_j)U_j'  , \sum_{k=1}^N w_{jk}*\left[ f'(U_k) U_k' \right] \cdot \left[  \Delta_k  - \Delta_j \right] \right\rangle.   \label{instareigs}
\end{align}
It can be shown Eq.~(\ref{instareigs}) is immediately satisfied for $j=1,...,M$, since $w_{jk}(x) \equiv 0$ for $k \neq j$. The remaining ($N-M$)-dimensional system for $(\Delta_{M+1},...,\Delta_N)$ can then be solved algebraically. As the independent phases $(\Delta_1,...,\Delta_M)$ determine the dynamics, we ignore the local impact of noise in the non-independent layers $(\Delta_{M+1},...,\Delta_N)$. We now demonstrate this calculation for a 3-layer model with two independent layers ($j=1,2$).

{\em Three-layer imploding star.} We begin by assuming the constituent functions take the form $w_{jj}(x) = \cos (x)$ ($j=1,..,3$); $h_j(x) = \sigma \cos (n_jx)$ ($j=1,2$); $w_{3j}(x) = \bar{w}_c (1+ \cos (x))/2$ ($j=1,2$); and $C_{jj}(x) = \pi \cos (x)$. In this case, $q_j(\Delta_j) = - {\mc C}_j \sin (n_j \Delta_j)/ (2 \sin (a))$ and $D_{jj} = \pi \ve^2/ (4 \sin^2(a))$ ($j=1,2$) in Eq.~(\ref{delsdej}), with ${\mc C}_j$ defined as in Eq.~(\ref{Cjform}). Note that $a_1 = a_2 = a$, but $a_3 \neq a$, due to synaptic input from layers 1 and 2. Thus, using Eq.~(\ref{instareigs}), we can solve to find that $\Delta_3(t) = (\Delta_1(t) + \Delta_2(t))/2$, so $\langle \Delta_3^2(t) \rangle = \left[ \langle \Delta_1^2(t) \rangle  + \langle \Delta_2^2(t) \rangle \right]/4$. Both the low-dimensional approximation, Eq.~(\ref{delsdej}), and the resulting variances compare well with our results from numerical simulation (Fig. \ref{fig9}). Also, even though there is no recurrence in this network, the fact that the output layer 3 receives two independent feedforward inputs means its estimate will be a weighted average of layers 1 and 2. Ultimately, this leads to more robust storage of the initial condition of the network in the output layer.

\section{Discussion}
\label{sec:discuss}

We have carried out a detailed analysis of the stochastic dynamics of bumps in multilayer neural fields. Importantly, the model incorporated both spatial heterogeneities and velocity inputs, to understand how these network features interacted with noise. In the absence of velocity input, we have shown that a bump's response to perturbations is shaped by the graph of the interlaminar architecture. Bumps in layers of the network that receive no feedback from other layers will not be affected by perturbations to the rest of the network. This lack of feedback to independent layers means that such feedforward networks are less robust to noise perturbations, since noise cancelation relies upon the presence of recurrent architecture~\cite{kilpatrick13b}. Recurrently coupled networks are more robust to noise perturbations, especially when layers possess heterogeneity. The most severe heterogeneities will tend to determine the stability of the entire network's bump solution in the presence of noise. However, the stabilizing effect of heterogeneities is disruptive to velocity integration, since it slows the propagation of velocity-driven bumps. Interestingly, noise can restore the propagation of bumps, so they move at a speed close to the input. We also extended this analysis to the case of networks with multiple independent layers, showing multiple phase variables are needed to describe each independent layer's bump. The non-independent layers are entrained by the phases of the independent layers. Our work extends previous results on the impact of noise~\cite{bressloff12}, heterogeneity~\cite{bressloff01}, and velocity input~\cite{zhang96} on the dynamics of continuum neural fields, to address how multilayer architectures shape networks' processing of spatially-relevant inputs.

Our multilayer network analysis need not be limited to layers that support stationary bump attractors. In particular, we expect that similar analyses could be performed on neural fields that support traveling waves~\cite{pinto01} or Turing patterns~\cite{coombes05}. It would be interesting to examine layers that individually support Turing patterns with different dominant frequencies, to see how interlaminar coupling  impacts the onset of pattern-formation and the frequency of the emerging pattern. We are also interested in extending this framework to multilayer networks whose individual layers support different classes of solution. For example, we could consider a network comprised of two layers wherein one layer supports bump attractors and the other supports stationary front solutions. In the case of excitatory feedforward input from the bump to the front layer, the front would expand only in response to the motion of the bump. Such a network could provide robust storage of visited locations during memory-guided visual search~\cite{chelazzi98} or spatial navigation~\cite{geva15}.

\appendix
\section{From a double-ring to a single layer velocity integration network} 
\label{dringone}
In Eq.~(\ref{nfmodel}) and in previous work~\cite{poll16}, we present a model with a spatially asymmetric weight function whose amplitude represents velocity input. Varying this input leads to a proportional rise in the velocity of moving bumps generated in the corresponding network. This single-layer network is a linear reduction of a ``double-ring" network, analyzed in detail in \cite{xie02}. Originally developed as a model of the head-direction system, the rings of the double-ring network each prefer either rightward or leftward velocity inputs. However, similar network architectures have been used to model the dynamics of activity in the brain's spatial navigation system~\cite{burak09}, as we consider here. We now demonstrate a reduction of the double-ring network to a single-ring network where inputs are given as a pre-factor to an integral term with asymmetric coupling, as in \cite{poll16}. In Section \ref{dringmulti}, we show how this reduction extends to a two-layer network, where each layer is a reduction of a ``double-ring."

We consider a slight variation on the model used in \cite{xie02}, so the nonlinearity filtering synaptic input is within, rather than outside, the convolution integrals. Note, it is typically possible to perform a mapping between such models~\cite{bressloff12}. In a double-ring model, there are two synaptic input variables $u_L(x,t)$ and $u_R(x,t)$, for leftward and rightward preferring velocity populations respectively, subject to the evolution equations
\begin{subequations} \label{dublring}
\begin{align}
\frac{\pd u_L}{\pd t} &= - u_L + w(x+\phi)*f(u_L(x,t)) + w(x-\phi)*f(u_R(x,t))+ I(t), \\
\frac{\pd u_R}{\pd t} &= -u_R + w(x+\phi)*f(u_L(x,t)) + w(x-\phi)*f(u_R(x,t)) - I(t),
\end{align}
\end{subequations}
where the nonzero shift $\phi > 0$ in either weight function $w(x\pm \phi)$ is crucial for generating traveling bumps in the input driven system ($I(t) \neq 0$). Note that the function $w(x) = w(-x)$ is a typical even-symmetric, lateral-inhibitory weight kernel, as described in Section \ref{sec:model}. Symmetric, stationary bump solutions $u_{L,R}(x,t) = U(x)$ to Eq.~(\ref{dublring}) are given by the equation:
\begin{align}
U(x) &= \bar{w}(x)*f(U(x)), \hspace{5mm} \bar{w}(x) = w(x+\phi) + w(x- \phi),  \label{dbeqn}
\end{align}
where $\bar{w}(x)$ is an even symmetric function, since $\bar{w}(-x) = w(-x+\phi) + w(-x- \phi) = w(x-\phi) + w(x+\phi) = \bar{w}(x)$. Input $I \neq 0$ is converted to bump velocity, which can be demonstrated by assuming $|I(t)| \ll 1$ and linearizing Eq.~(\ref{dublring}) using the ansatz, $u_j(x,t) = U(x - X(t)-\psi_j)+\ve \Phi_j(x,t) + {\mc O}(\ve^2)$ ($j=L,R$): 
\begin{align}
\frac{\pd}{\pd t} \left( \begin{array}{c} \Phi_L (x,t) \\ \Phi_R(x,t) \end{array} \right) = {\mc L} \left( \begin{array}{c} \Phi_L (x,t) \\ \Phi_R(x,t) \end{array} \right) + \left( \begin{array}{c} \ve v(t) U'(x) + I(t) \\ \ve v(t) U'(x) - I(t) \end{array} \right),  \label{dringlin}
\end{align}
where $\ve v(t) = \dot{X}(t)$ is the bump's velocity, and the linear operator
\begin{align*}
 {\mc L} \left( \begin{array}{c} \Phi_L \\ \Phi_R \end{array} \right) = \left( \begin{array}{c} - \Phi_L + w(x+\phi)*[f'(U)\Phi_L] + w(x-\phi)*[f'(U)\Phi_R]  \\  - \Phi_R + w(x+\phi)*[f'(U)\Phi_L] + w(x-\phi)*[f'(U)\Phi_R]   \end{array} \right).
\end{align*}
For solutions to Eq.~(\ref{dringlin}) to be bounded, we require the inhomogeneous portion to be orthogonal to the nullspace of the adjoint linear operator, defined as
\begin{align*}
 {\mc L}^* \left( \begin{array}{c} \Psi_L \\ \Psi_R \end{array} \right) = \left( \begin{array}{c} - \Psi_L + f'(U)\cdot w(x+\phi)*[\Psi_L+ \Psi_R]  \\  - \Psi_R + f'(U) \cdot w(x-\phi)*[\Psi_L + \Psi_R]   \end{array} \right) \equiv \left( \begin{array}{c} 0 \\ 0 \end{array} \right).
\end{align*}
This leads to the following equation for the dependence of the bumps' velocity $\ve v(t)$ on the input $I(t)$:
\begin{align*}
\ve v(t) = \frac{\langle I(t), \Psi_L(x) - \Psi_R(x) \rangle }{\langle U'(x), \Psi_L(x) + \Psi_R(x) \rangle}. 
\end{align*}
Thus, there is a proportional increase in the velocity $\ve v(t)$ corresponding to an increase in the input $I(t)$, to linear order. By differentiating Eq.~(\ref{dbeqn}), we see a solution to Eq.~(\ref{dringlin}) is $\Phi_{L,R}(x,t) = U'(x)$. Thus, up to ${\mc O}(\ve^2)$, we can approximate $u_{L,R}(x,t) \approx U(x - X(t) - \psi_{L,R})$. Dropping subscripts on the $u_j$ ($j=L,R$) formulae and differentiating with respect to $t$, we find
\begin{align}
\frac{\pd u(x,t)}{\pd t} = - \ve v(t) U'(x-X(t) + \psi),  \label{drtder}
\end{align}
and we can further incorporate the formula for the stationary bump by replacing $U(x-X(t) - \psi)$ with $u(x,t)$ in Eq.~(\ref{dbeqn}), and adding the equation to Eq.~(\ref{drtder}). Subsequently, a differentiation of that formula, with $U(x-X(t) - \psi)$ replaced with $u(x,t)$ means the $U'(x-X(t)-\psi)$ in Eq.~(\ref{drtder}) can also be replaced to yield
\begin{align}
\frac{\pd u(x,t)}{\pd t} = - u + \bar{w}*f(u) - \ve v(t) \left[\bar{w}'\right]*f(u),  \label{velueqn}
\end{align}
so Eq.~(\ref{velueqn}) describes the dynamics of Eq.~(\ref{dublring}) to linear order in $I(t)$.

\section{Reduction in a multilayer velocity integration network}
\label{dringmulti}
The double-ring model, Eq.~(\ref{dublring}), can be extended to the case of two layers (of double-rings), each receiving independent velocity-producing inputs. Now, there are four synaptic input variables ($u_{L1}, u_{R1},u_{L2},u_{R2}$), where $u_{jk}$ corresponds to the variable in the $k$th layer preferring $j$ ($L:$ left or $R:$ right)-ward velocity. These are subject to the evolution equations
\begin{align}
\dot{u_{L1}} &= -u_{L1} + w(x+\phi)*\left[ f(u_{L1}) + \alpha_c f(u_{L2}) \right] + w(x-\phi)*\left[ f(u_{R1}) + \alpha_c f(u_{R2}) \right] + I(t), \label{twodubring} \\
\dot{u_{R1}} &= -u_{R1} + w(x+\phi)*\left[ f(u_{L1}) + \alpha_c f(u_{L2}) \right] + w(x-\phi)*\left[ f(u_{R1}) + \alpha_c f(u_{R2}) \right] - I(t), \nonumber \\
\dot{u_{L2}} &= -u_{L2} + w(x+\phi)*\left[ f(u_{L2}) + \alpha_c f(u_{L1}) \right] + w(x-\phi)*\left[ f(u_{R2}) + \alpha_c f(u_{R1}) \right] + I(t), \nonumber \\
\dot{u_{R2}} &= -u_{R2} + w(x+\phi)*\left[ f(u_{L2}) + \alpha_c f(u_{L1}) \right] + w(x-\phi)*\left[ f(u_{R2}) + \alpha_c f(u_{R1}) \right] - I(t), \nonumber
\end{align}
where again the nonzero shift $\phi>0$ causes bumps to travel when $I(t) \neq 0$, and $\alpha_c$ represents the coupling between layers 1 and 2. Since $w(x) = w(-x)$ is even, there are symmetric, stationary bump solutions $u_{jk}(x,t) = U(x)$ ($j=L,R$, $k=1,2$) to Eq.~(\ref{twodubring}), given by:
\begin{align}
U(x) = (1+ \alpha_c) \bar{w}(x)*f(U(x)), \hspace{5mm} \bar{w}(x) = w(x+ \phi) + w(x-\phi), \label{twobumplin}
\end{align}
and note $\bar{w}(x)$ is even. When $|I(t)| \ll 1$, we linearize Eq.~(\ref{twodubring}) assuming $u_{jk}(x,t) = U(x- X(t) - \psi_{jk}) + \ve \Phi_{jk}(x,t) + {\mc O}(\ve^2)$ for $j=L,R$ and $k=1,2$:
\begin{align}
\left( \frac{\pd}{\pd t} - {\mc L} \right) \left( \begin{array}{c} \Phi_{L1}(x,t) \\ \Phi_{R1}(x,t) \\ \Phi_{L2}(x,t) \\ \Phi_{R2}(x,t) \end{array} \right) = \left( \begin{array}{c} \ve v(t)U'(x) + I(t) \\ \ve v(t)U'(x) - I(t) \\ \ve v(t)U'(x) + I(t) \\ \ve v(t)U'(x) - I(t) \end{array} \right), \label{tworinglin}
\end{align}
where $\ve v(t) = \dot{X}(t)$ is the bumps' velocity, and the linear operator
\begin{align*}
{\mc L} \left( \begin{array}{c} \Phi_{L1} \\ \Phi_{R1} \\ \Phi_{L2} \\ \Phi_{R2} \end{array} \right) = \left( \begin{array}{c} - \Phi_{L1} + w(x+ \phi)*\left[f'(U) \left(\Phi_{L1}+\alpha_c\Phi_{L2}\right)\right] + w(x- \phi)*\left[f'(U) \left(\Phi_{R1}+\alpha_c\Phi_{R2}\right)\right] \\
- \Phi_{R1} + w(x+ \phi)*\left[f'(U) \left(\Phi_{L1}+\alpha_c\Phi_{L2}\right)\right] + w(x- \phi)*\left[f'(U) \left(\Phi_{R1}+\alpha_c\Phi_{R2}\right)\right] \\
- \Phi_{L2} + w(x+ \phi)*\left[f'(U) \left(\Phi_{L2}+\alpha_c\Phi_{L1}\right)\right] + w(x- \phi)*\left[f'(U) \left(\Phi_{R2}+\alpha_c\Phi_{R1}\right)\right] \\
- \Phi_{R2} + w(x+ \phi)*\left[f'(U) \left(\Phi_{L2}+\alpha_c\Phi_{L1}\right)\right] + w(x- \phi)*\left[f'(U) \left(\Phi_{R2}+\alpha_c\Phi_{R1}\right)\right] \end{array} \right).
\end{align*}
For solutions to Eq.~(\ref{tworinglin}) to be bounded, we require the right hand side to be orthogonal to the nullspace of ${\mc L}^*$, defined:
\begin{align*}
{\mc L}^* \left( \begin{array}{c} \Psi_{L1} \\ \Psi_{R1} \\ \Psi_{L2} \\ \Psi_{R2} \end{array} \right) &= \left( \begin{array}{c} - \Psi_{L1} + f'(U) \cdot w(x+\phi)*\left[\Psi_{L1} + \Psi_{R1} + \alpha_c \left(\Psi_{L2}+\Psi_{R2} \right) \right] \\ - \Psi_{R1} + f'(U) \cdot w(x-\phi)*\left[\Psi_{L1} + \Psi_{R1} + \alpha_c \left(\Psi_{L2}+\Psi_{R2} \right) \right] \\ - \Psi_{L2} + f'(U) \cdot w(x+\phi)*\left[\Psi_{L2} + \Psi_{R2} + \alpha_c \left(\Psi_{L1}+\Psi_{R1} \right) \right] \\ - \Psi_{R2} + f'(U) \cdot w(x-\phi)*\left[\Psi_{L2} + \Psi_{R2} + \alpha_c \left(\Psi_{L1}+\Psi_{R1} \right) \right]  \end{array} \right) \equiv 0.
\end{align*}
This leads to the following linear equation, relating $\ve v(t)$ to $I(t)$:
\begin{align*}
\ve v(t) = \frac{\langle I(t) ,  \Psi_{L1}(x)+\Psi_{L2}(x) - \Psi_{R1}(x)-\Psi_{R2}(x) \rangle}{\langle U'(x), \Psi_{L1}(x)+\Psi_{L2}(x) + \Psi_{R1}(x)+\Psi_{R2}(x) \rangle}.
\end{align*}
Lastly, noting $\Phi_{jk}(x,t) = U'(x)$ solves Eq.~(\ref{tworinglin}), we can approximate $u_{jk} \approx U(x-X(t)-\psi_{jk})$ ($j=L,R$ and $k=1,2$) up to ${\mc O}(\ve^2)$. Dropping the subscripts and differentiating with respect to $t$, we again find Eq.~(\ref{drtder}). Next, replacing $U(x-X(t)-\psi)$ with $u(x,t)$ in Eq.~(\ref{twobumplin}) and adding to Eq.~(\ref{drtder}) as well as plugging this equation in for $U'(x-X(t)-\psi)$ yields
\begin{align}
\frac{\pd u(x,t)}{\pd t} = -u + (1+ \alpha_c) \bar{w}*f(u) - \ve v(t) (1+ \alpha_c) \left[\bar{w}' \right]*f(u).  \label{twodring2}
\end{align}
Note, in the case of asymmetric coupling between either double ring, we would expect two distinct forms of Eq.~(\ref{twodring2}), where $\bar{w}'$ was different for either. This full asymmetry for an arbitrary number of layers is captured by the asymmetric weight functions given by Eqs.~(\ref{nfmodel}) and (\ref{velwt}).

\bibliographystyle{siamplain}

\begin{thebibliography}{10}

\bibitem{aksay01}
{\sc E.~Aksay, G.~Gamkrelidze, H.~Seung, R.~Baker, and D.~Tank}, {\em In vivo
  intracellular recording and perturbation of persistent activity in a neural
  integrator}, Nature neuroscience, 4 (2001), pp.~184--193.

\bibitem{amari77}
{\sc S.-i. Amari}, {\em Dynamics of pattern formation in lateral-inhibition
  type neural fields}, Biological cybernetics, 27 (1977), pp.~77--87.

\bibitem{anderson03}
{\sc M.~I. Anderson and K.~J. Jeffery}, {\em Heterogeneous modulation of place
  cell firing by changes in context}, The Journal of Neuroscience, 23 (2003),
  pp.~8827--8835.

\bibitem{anello04}
{\sc G.~Anello and G.~Cordaro}, {\em Existence of solutions and bifurcation
  points to hammerstein equations with essentially bounded kernel}, Journal of
  mathematical analysis and applications, 298 (2004), pp.~292--297.

\bibitem{atkinson76}
{\sc K.~Atkinson}, {\em A survey of numerical methods for the solution of
  Fredholm integral equations of the second kind}, SIAM, 1976.

\bibitem{baddeley03}
{\sc A.~Baddeley}, {\em Working memory: looking back and looking forward},
  Nature reviews neuroscience, 4 (2003), pp.~829--839.

\bibitem{battaglia04}
{\sc F.~P. Battaglia, G.~R. Sutherland, and B.~L. McNaughton}, {\em Local
  sensory cues and place cell directionality: additional evidence of
  prospective coding in the hippocampus}, The Journal of Neuroscience, 24
  (2004), pp.~4541--4550.

\bibitem{bays15}
{\sc P.~M. Bays}, {\em Spikes not slots: noise in neural populations limits
  working memory}, Trends in cognitive sciences, 19 (2015), pp.~431--438.

\bibitem{bressloff01}
{\sc P.~C. Bressloff}, {\em Traveling fronts and wave propagation failure in an
  inhomogeneous neural network}, Physica D: Nonlinear Phenomena, 155 (2001),
  pp.~83--100.

\bibitem{bressloff12}
{\sc P.~C. Bressloff}, {\em Spatiotemporal dynamics of continuum neural
  fields}, Journal of Physics A: Mathematical and Theoretical, 45 (2011),
  p.~033001.

\bibitem{bressloff15}
{\sc P.~C. Bressloff and Z.~P. Kilpatrick}, {\em Nonlinear langevin equations
  for wandering patterns in stochastic neural fields}, SIAM Journal on Applied
  Dynamical Systems, 14 (2015), pp.~305--334.

\bibitem{brody03}
{\sc C.~D. Brody, R.~Romo, and A.~Kepecs}, {\em Basic mechanisms for graded
  persistent activity: discrete attractors, continuous attractors, and dynamic
  representations}, Current opinion in neurobiology, 13 (2003), pp.~204--211.

\bibitem{brunel99}
{\sc N.~Brunel and V.~Hakim}, {\em Fast global oscillations in networks of
  integrate-and-fire neurons with low firing rates}, Neural computation, 11
  (1999), pp.~1621--1671.

\bibitem{burak08}
{\sc Y.~Burak and I.~R. Fiete}, {\em Grid cells: the position code, neural
  network models of activity, and the problem of learning}, Hippocampus, 18
  (2008), pp.~1283--1300.

\bibitem{burak09}
{\sc Y.~Burak and I.~R. Fiete}, {\em Accurate path integration in continuous
  attractor network models of grid cells}, PLoS Comput Biol, 5 (2009),
  p.~e1000291.

\bibitem{buzsaki13}
{\sc G.~Buzs{\'a}ki and E.~I. Moser}, {\em Memory, navigation and theta rhythm
  in the hippocampal-entorhinal system}, Nature neuroscience, 16 (2013),
  pp.~130--138.

\bibitem{carroll14}
{\sc S.~Carroll, K.~Josi{\'c}, and Z.~P. Kilpatrick}, {\em Encoding certainty
  in bump attractors}, Journal of computational neuroscience, 37 (2014),
  pp.~29--48.

\bibitem{chelazzi98}
{\sc L.~Chelazzi, J.~Duncan, E.~K. Miller, and R.~Desimone}, {\em Responses of
  neurons in inferior temporal cortex during memory-guided visual search},
  Journal of neurophysiology, 80 (1998), pp.~2918--2940.

\bibitem{compte00}
{\sc A.~Compte, N.~Brunel, P.~S. Goldman-Rakic, and X.-J. Wang}, {\em Synaptic
  mechanisms and network dynamics underlying spatial working memory in a
  cortical network model}, Cerebral Cortex, 10 (2000), pp.~910--923.

\bibitem{constantinidis04}
{\sc C.~Constantinidis and X.-J. Wang}, {\em A neural circuit basis for spatial
  working memory}, The Neuroscientist, 10 (2004), pp.~553--565.

\bibitem{coombes05}
{\sc S.~Coombes}, {\em Waves, bumps, and patterns in neural field theories},
  Biological cybernetics, 93 (2005), pp.~91--108.

\bibitem{coombes04}
{\sc S.~Coombes and M.~R. Owen}, {\em Evans functions for integral neural field
  equations with heaviside firing rate function}, SIAM Journal on Applied
  Dynamical Systems, 3 (2004), pp.~574--600.

\bibitem{curtis06}
{\sc C.~Curtis}, {\em Prefrontal and parietal contributions to spatial working
  memory}, Neuroscience, 139 (2006), pp.~173--180.

\bibitem{durstewitz00}
{\sc D.~Durstewitz, J.~K. Seamans, and T.~J. Sejnowski}, {\em
  Neurocomputational models of working memory}, Nature neuroscience, 3 (2000),
  pp.~1184--1191.

\bibitem{ermentrout98}
{\sc B.~Ermentrout}, {\em Neural networks as spatio-temporal pattern-forming
  systems}, Reports on progress in physics, 61 (1998), p.~353.

\bibitem{faisal08}
{\sc A.~A. Faisal, L.~P. Selen, and D.~M. Wolpert}, {\em Noise in the nervous
  system}, Nature reviews neuroscience, 9 (2008), pp.~292--303.

\bibitem{folias11}
{\sc S.~Folias and G.~Ermentrout}, {\em New patterns of activity in a pair of
  interacting excitatory-inhibitory neural fields}, Physical review letters,
  107 (2011), p.~228103.

\bibitem{folias04}
{\sc S.~E. Folias and P.~C. Bressloff}, {\em Breathing pulses in an excitatory
  neural network}, SIAM Journal on Applied Dynamical Systems, 3 (2004),
  pp.~378--407.

\bibitem{folias12}
{\sc S.~E. Folias and G.~B. Ermentrout}, {\em Bifurcations of stationary
  solutions in an interacting pair of ei neural fields}, SIAM Journal on
  Applied Dynamical Systems, 11 (2012), pp.~895--938.

\bibitem{funahashi89}
{\sc S.~Funahashi, C.~J. Bruce, and P.~S. Goldman-Rakic}, {\em Mnemonic coding
  of visual space in the monkey's dorsolateral prefrontal cortex}, Journal of
  neurophysiology, 61 (1989), pp.~331--349.

\bibitem{geva15}
{\sc M.~Geva-Sagiv, L.~Las, Y.~Yovel, and N.~Ulanovsky}, {\em Spatial cognition
  in bats and rats: from sensory acquisition to multiscale maps and
  navigation}, Nature Reviews Neuroscience, 16 (2015), pp.~94--108.

\bibitem{goldmanrakic95}
{\sc P.~S. Goldman-Rakic}, {\em Cellular basis of working memory}, Neuron, 14
  (1995), pp.~477--485.

\bibitem{guo05}
{\sc Y.~Guo and C.~C. Chow}, {\em Existence and stability of standing pulses in
  neural networks: I. existence}, SIAM Journal on Applied Dynamical Systems, 4
  (2005), pp.~217--248.

\bibitem{hafting05}
{\sc T.~Hafting, M.~Fyhn, S.~Molden, M.-B. Moser, and E.~I. Moser}, {\em
  Microstructure of a spatial map in the entorhinal cortex}, Nature, 436
  (2005), pp.~801--806.

\bibitem{hammerstein30}
{\sc A.~Hammerstein}, {\em Nichtlineare integralgleichungen nebst anwendungen},
  Acta Mathematica, 54 (1930), pp.~117--176.

\bibitem{hansel98}
{\sc D.~Hansel and H.~Sompolinsky}, {\em Modeling feature selectivity in local
  cortical circuits}, in Methods in neuronal modeling: From ions to networks,
  C.~Koch and I.~Segev, eds., Cambridge: MIT, 1998, ch.~13, pp.~499--567.

\bibitem{haxby00}
{\sc J.~V. Haxby, L.~Petit, L.~G. Ungerleider, and S.~M. Courtney}, {\em
  Distinguishing the functional roles of multiple regions in distributed neural
  systems for visual working memory}, Neuroimage, 11 (2000), pp.~145--156.

\bibitem{kastner07}
{\sc S.~Kastner, K.~DeSimone, C.~S. Konen, S.~M. Szczepanski, K.~S. Weiner, and
  K.~A. Schneider}, {\em Topographic maps in human frontal cortex revealed in
  memory-guided saccade and spatial working-memory tasks}, Journal of
  neurophysiology, 97 (2007), pp.~3494--3507.

\bibitem{kilpatrick13b}
{\sc Z.~P. Kilpatrick}, {\em Interareal coupling reduces encoding variability
  in multi-area models of spatial working memory}, Frontiers in computational
  neuroscience, 7 (2013), p.~82.

\bibitem{kilpatrick15}
{\sc Z.~P. Kilpatrick}, {\em Delay stabilizes stochastic motion of bumps in
  layered neural fields}, Physica D: Nonlinear Phenomena, 295 (2015),
  pp.~30--45.

\bibitem{kilpatrick13}
{\sc Z.~P. Kilpatrick and B.~Ermentrout}, {\em Wandering bumps in stochastic
  neural fields}, SIAM Journal on Applied Dynamical Systems, 12 (2013),
  pp.~61--94.

\bibitem{kilpatrick13c}
{\sc Z.~P. Kilpatrick, B.~Ermentrout, and B.~Doiron}, {\em Optimizing working
  memory with heterogeneity of recurrent cortical excitation}, The Journal of
  Neuroscience, 33 (2013), pp.~18999--19011.

\bibitem{knierim12}
{\sc J.~J. Knierim and K.~Zhang}, {\em Attractor dynamics of spatially
  correlated neural activity in the limbic system}, Annual review of
  neuroscience, 35 (2012), pp.~267--285.

\bibitem{ko11}
{\sc H.~Ko, S.~B. Hofer, B.~Pichler, K.~A. Buchanan, P.~J. Sj{\"o}str{\"o}m,
  and T.~D. Mrsic-Flogel}, {\em Functional specificity of local synaptic
  connections in neocortical networks}, Nature, 473 (2011), pp.~87--91.

\bibitem{laing01}
{\sc C.~R. Laing and C.~C. Chow}, {\em Stationary bumps in networks of spiking
  neurons}, Neural Computation, 13 (2001), pp.~1473--1494.

\bibitem{laing01b}
{\sc C.~R. Laing and A.~Longtin}, {\em Noise-induced stabilization of bumps in
  systems with long-range spatial coupling}, Physica D, 160 (2001), pp.~149 --
  172.

\bibitem{laing02}
{\sc C.~R. Laing, W.~C. Troy, B.~Gutkin, and G.~B. Ermentrout}, {\em Multiple
  bumps in a neuronal model of working memory}, SIAM Journal on Applied
  Mathematics, 63 (2002), pp.~62--97.

\bibitem{lee04}
{\sc I.~Lee, D.~Yoganarasimha, G.~Rao, and J.~J. Knierim}, {\em Comparison of
  population coherence of place cells in hippocampal subfields ca1 and ca3},
  Nature, 430 (2004), pp.~456--459.

\bibitem{lindner01}
{\sc B.~Lindner, M.~Kostur, and L.~Schimansky-Geier}, {\em Optimal diffusive
  transport in a tilted periodic potential}, Fluctuation and Noise Letters, 1
  (2001), pp.~R25--R39.

\bibitem{longtin93}
{\sc A.~Longtin}, {\em Stochastic resonance in neuron models}, Journal of
  statistical physics, 70 (1993), pp.~309--327.

\bibitem{mcnaughton06}
{\sc B.~L. McNaughton, F.~P. Battaglia, O.~Jensen, E.~I. Moser, and M.-B.
  Moser}, {\em Path integration and the neural basis of the'cognitive map'},
  Nature Reviews Neuroscience, 7 (2006), pp.~663--678.

\bibitem{moser08}
{\sc E.~I. Moser, E.~Kropff, and M.-B. Moser}, {\em Place cells, grid cells,
  and the brain's spatial representation system}, Annual Review of
  Neuroscience, 31 (2008), pp.~69--89.

\bibitem{obrien90}
{\sc W.~J. O'brien, H.~I. Browman, and B.~I. Evans}, {\em Search strategies of
  foraging animals}, American Scientist, 78 (1990), pp.~152--160.

\bibitem{pesaran02}
{\sc B.~Pesaran, J.~S. Pezaris, M.~Sahani, P.~P. Mitra, and R.~A. Andersen},
  {\em Temporal structure in neuronal activity during working memory in macaque
  parietal cortex}, Nature neuroscience, 5 (2002), pp.~805--811.

\bibitem{pfeiffer15}
{\sc B.~E. Pfeiffer and D.~J. Foster}, {\em Autoassociative dynamics in the
  generation of sequences of hippocampal place cells}, Science, 349 (2015),
  pp.~180--183.

\bibitem{pinto01}
{\sc D.~J. Pinto and G.~B. Ermentrout}, {\em Spatially structured activity in
  synaptically coupled neuronal networks: I. traveling fronts and pulses}, SIAM
  journal on Applied Mathematics, 62 (2001), pp.~206--225.

\bibitem{poll15}
{\sc D.~Poll and Z.~P. Kilpatrick}, {\em Stochastic motion of bumps in planar
  neural fields}, SIAM Journal on Applied Mathematics, 75 (2015),
  pp.~1553--1577.

\bibitem{poll16}
{\sc D.~B. Poll, K.~Nguyen, and Z.~P. Kilpatrick}, {\em Sensory feedback in a
  bump attractor model of path integration}, Journal of computational
  neuroscience, 40 (2016), pp.~137--155.

\bibitem{qi11}
{\sc X.-L. Qi, T.~Meyer, T.~R. Stanford, and C.~Constantinidis}, {\em Changes
  in prefrontal neuronal activity after learning to perform a spatial working
  memory task}, Cerebral Cortex,  (2011).

\bibitem{rao99}
{\sc S.~G. Rao, G.~V. Williams, and P.~S. Goldman-Rakic}, {\em Isodirectional
  tuning of adjacent interneurons and pyramidal cells during working memory:
  evidence for microcolumnar organization in pfc}, Journal of Neurophysiology,
  81 (1999), pp.~1903--1916.

\bibitem{renart04}
{\sc A.~Renart, N.~Brunel, and X.-J. Wang}, {\em Mean-field theory of
  irregularly spiking neuronal populations and working memory in recurrent
  cortical networks}, Computational neuroscience: A comprehensive approach,
  (2004), pp.~431--490.

\bibitem{renart03}
{\sc A.~Renart, P.~Song, and X.-J. Wang}, {\em Robust spatial working memory
  through homeostatic synaptic scaling in heterogeneous cortical networks},
  Neuron, 38 (2003), pp.~473--485.

\bibitem{risken84}
{\sc H.~Risken}, {\em Fokker-planck equation}, in The Fokker-Planck Equation,
  Springer, 1984, pp.~63--95.

\bibitem{rowe00}
{\sc J.~B. Rowe, I.~Toni, O.~Josephs, R.~S. Frackowiak, and R.~E. Passingham},
  {\em The prefrontal cortex: response selection or maintenance within working
  memory?}, Science, 288 (2000), pp.~1656--1660.

\bibitem{samsonovich97}
{\sc A.~Samsonovich and B.~L. McNaughton}, {\em Path integration and cognitive
  mapping in a continuous attractor neural network model}, The Journal of
  neuroscience, 17 (1997), pp.~5900--5920.

\bibitem{sandstede02}
{\sc B.~Sandstede}, {\em Stability of travelling waves}, Handbook of dynamical
  systems, 2 (2002), pp.~983--1055.

\bibitem{sargolini06}
{\sc F.~Sargolini, M.~Fyhn, T.~Hafting, B.~L. McNaughton, M.~P. Witter, M.-B.
  Moser, and E.~I. Moser}, {\em Conjunctive representation of position,
  direction, and velocity in entorhinal cortex}, Science, 312 (2006),
  pp.~758--762.

\bibitem{schluppeck06}
{\sc D.~Schluppeck, C.~E. Curtis, P.~W. Glimcher, and D.~J. Heeger}, {\em
  Sustained activity in topographic areas of human posterior parietal cortex
  during memory-guided saccades}, The Journal of neuroscience, 26 (2006),
  pp.~5098--5108.

\bibitem{schneidman03}
{\sc E.~Schneidman, W.~Bialek, and M.~J. Berry}, {\em Synergy, redundancy, and
  independence in population codes}, the Journal of Neuroscience, 23 (2003),
  pp.~11539--11553.

\bibitem{seung96}
{\sc H.~S. Seung}, {\em How the brain keeps the eyes still}, Proceedings of the
  National Academy of Sciences, 93 (1996), pp.~13339--13344.

\bibitem{silvester00}
{\sc J.~R. Silvester}, {\em Determinants of block matrices}, The Mathematical
  Gazette, 84 (2000), pp.~460--467.

\bibitem{taube07}
{\sc J.~S. Taube}, {\em The head direction signal: origins and sensory-motor
  integration}, Annu. Rev. Neurosci., 30 (2007), pp.~181--207.

\bibitem{veltz10}
{\sc R.~Veltz and O.~Faugeras}, {\em Local/global analysis of the stationary
  solutions of some neural field equations}, SIAM Journal on Applied Dynamical
  Systems, 9 (2010), pp.~954--998.

\bibitem{wang99}
{\sc X.-J. Wang}, {\em Synaptic basis of cortical persistent activity: the
  importance of nmda receptors to working memory}, The Journal of Neuroscience,
  19 (1999), pp.~9587--9603.

\bibitem{wang06}
{\sc Y.~Wang, H.~Markram, P.~H. Goodman, T.~K. Berger, J.~Ma, and P.~S.
  Goldman-Rakic}, {\em Heterogeneity in the pyramidal network of the medial
  prefrontal cortex}, Nature neuroscience, 9 (2006), pp.~534--542.

\bibitem{wills05}
{\sc T.~J. Wills, C.~Lever, F.~Cacucci, N.~Burgess, and J.~O'Keefe}, {\em
  Attractor dynamics in the hippocampal representation of the local
  environment}, Science, 308 (2005), pp.~873--876.

\bibitem{wimmer14}
{\sc K.~Wimmer, D.~Q. Nykamp, C.~Constantinidis, and A.~Compte}, {\em Bump
  attractor dynamics in prefrontal cortex explains behavioral precision in
  spatial working memory}, Nature neuroscience, 17 (2014), pp.~431--439.

\bibitem{xie02}
{\sc X.~Xie, R.~H. Hahnloser, and H.~S. Seung}, {\em Double-ring network model
  of the head-direction system}, Physical Review E, 66 (2002), p.~041902.

\bibitem{yoon13}
{\sc K.~Yoon, M.~A. Buice, C.~Barry, R.~Hayman, N.~Burgess, and I.~R. Fiete},
  {\em Specific evidence of low-dimensional continuous attractor dynamics in
  grid cells}, Nature neuroscience, 16 (2013), pp.~1077--1084.

\bibitem{yoon16}
{\sc K.~Yoon, S.~Lewallen, A.~A. Kinkhabwala, D.~W. Tank, and I.~R. Fiete},
  {\em Grid cell responses in 1d environments assessed as slices through a 2d
  lattice}, Neuron, 89 (2016), pp.~1086--1099.

\bibitem{zhang96}
{\sc K.~Zhang}, {\em Representation of spatial orientation by the intrinsic
  dynamics of the head-direction cell ensemble: a theory}, The journal of
  neuroscience, 16 (1996), pp.~2112--2126.

\end{thebibliography}

\end{document}